\documentclass[12pt,preprint]{aastex}

\slugcomment{accepted by ApJS}

\shorttitle{SXDS Optical Catalogs}
\shortauthors{Furusaswa et al.}

\begin{document}


\title{
The Subaru/XMM-Newton Deep Survey (SXDS) - II. Optical Imaging and Photometric Catalogs
\footnote{Based on data collected at Subaru Telescope, which is operated
by the National Astronomical Observatory of Japan.}
}


\author{Hisanori Furusawa\altaffilmark{1},
George Kosugi\altaffilmark{2}, 
Masayuki Akiyama\altaffilmark{1}, 
Tadafumi Takata\altaffilmark{2}, 
Kazuhiro Sekiguchi\altaffilmark{2}, 
Ichi Tanaka\altaffilmark{1},
Ikuru Iwata\altaffilmark{3}, 
Masaru Kajisawa\altaffilmark{2}, 
Naoki Yasuda\altaffilmark{4},
Mamoru Doi\altaffilmark{5}, 
Masami Ouchi\altaffilmark{6,7},
Chris Simpson\altaffilmark{8},
Kazuhiro Shimasaku\altaffilmark{9,10},
Toru Yamada\altaffilmark{11},
Junko Furusawa\altaffilmark{1}, 
Tomoki Morokuma\altaffilmark{2},
Catherine M. Ishida\altaffilmark{1}, 
Kentaro Aoki\altaffilmark{1}, 
Tetsuharu Fuse\altaffilmark{1}, 
Masatoshi Imanishi\altaffilmark{2}, 
Masanori Iye\altaffilmark{2}, 
Hiroshi Karoji\altaffilmark{2},
Naoto Kobayashi\altaffilmark{5}, 
Tadayuki Kodama\altaffilmark{2}, 
Yutaka Komiyama\altaffilmark{1,2}, 
Yoshitomo Maeda\altaffilmark{12}, 
Satoshi Miyazaki\altaffilmark{1,2}, 
Yoshihiko Mizumoto\altaffilmark{2},
Fumiaki Nakata\altaffilmark{2}, 
Jun'ichi Noumaru\altaffilmark{1}, 
Ryusuke Ogasawara\altaffilmark{2},
Sadanori Okamura\altaffilmark{9,10}, 
Tomoki Saito\altaffilmark{13},
Toshiyuki Sasaki\altaffilmark{1},
Yoshihiro Ueda\altaffilmark{14},
and Michitoshi Yoshida\altaffilmark{3}
}

%
%
%


\altaffiltext{1}{Subaru Telescope, National Astronomical Observatory of Japan, %
 650 North A'ohoku Place, Hilo, HI 96720 USA; furusawa@subaru.naoj.org}
\altaffiltext{2}{National Astronomical Observatory of Japan, Mitaka, Tokyo 181-8588 Japan}
\altaffiltext{3}{Okayama Astrophysical Observatory, National
 Astronomical Observatory of Japan, 3037-5 Honjo, Kamogata, Asakuchi, Okayama
 719-0232 Japan}
\altaffiltext{4}{Institute for Cosmic Ray Research, University of Tokyo, 5-1-5 Kashiwa-no-Ha, Kashiwa, Chiba 277-8582 Japan}
\altaffiltext{5}{Institute of Astronomy, School of Science, %
 University of Tokyo, 2-21-1 Osawa, Mitaka, Tokyo 181-8588 Japan}
\altaffiltext{6}{Observatories of the Carnegie Institution of Washington, 813 Santa Barbara Street, Pasadena, CA 91101}
\altaffiltext{7}{Carnegie Fellow}
\altaffiltext{8}{Astrophysics Research Institute, Liverpool John Moores University, Twelve Quays House, Egerton Wharf, Birkenhead CH41 1LD UK}
\altaffiltext{9}{Department of Astronomy, School of Science, %
 University of Tokyo, 7-3-1 Hongo, Bunkyo, Tokyo 113-0033 Japan}
\altaffiltext{10}{Reseach Center for the Early Universe, School of Science, University of Tokyo, 7-3-1 Hongo, Bunkyo, Tokyo 113-0033 Japan}
\altaffiltext{11}{Astronomical Institute, Graduate School of Science,
   Tohoku University, Aramaki, Aoba, Sendai, Miyagi 980-8578 Japan}
\altaffiltext{12}{Institute of Space and Astronautical Science, Japan Aerospace Exploration Agency, Sagamihara, Kanagawa 229-8510 Japan}
\altaffiltext{13}{Physics Department, Graduate School of Science,
   Ehime University, 2-5 Bunkyou, Matsuyama,Ehime 790-8577 Japan}
\altaffiltext{14}{Department of Astronomy, Kyoto University, Sakyo-ku, Kyoto 606-8502 Japan}


\begin{abstract}
We present multi-waveband optical imaging data obtained from observations of
 the Subaru/XMM-Newton Deep Survey (SXDS). The survey field, centered at
 R.A. = $02^{h}18^{m}00^{s}$, decl. = $-05^{\circ}00'00''$, has been the
 focus of a wide range of multi-wavelength observing programs spanning
 from X-ray to radio wavelengths.
A large part of the optical imaging observations are carried out with Suprime-Cam on
 Subaru Telescope at Mauna Kea in the course of Subaru Telescope ``Observatory Projects''. 
This paper describes our optical observations, data reduction and
 analysis procedures employed, and the characteristics of the data products.
A total area of 1.22 deg$^{2}$ is covered in five contiguous sub-fields, 
each of which corresponds to a single Suprime-Cam field of view
 ($\sim34'\times 27'$ ), in five broad-band filters $B, V, R_c, i', z'$ to 
the depths of $B=28.4, V=27.8, R_c=27.7, i'=27.7$ and $z'=26.6$ ($AB, 3\,\sigma, \phi=2''$).
The data are reduced and compiled into five multi-waveband photometric
catalogs, separately for each Suprime-Cam pointing. The $i'$-band catalogs contain about 900,000 objects, making the SXDS catalogs one of the largest multi-waveband catalogs in
 corresponding depth and area coverage.
The SXDS catalogs can be used for an extensive range of astronomical
 applications such as the number density of the Galactic halo stars to the 
large scale structures at the distant universe. 
The number counts of galaxies are derived and compared with those of existing deep 
extragalactic surveys.
The optical data, the source catalogs, and configuration files used to
 create the catalogs are publicly available via the SXDS web page ({\tt
 http://www.naoj.org/Science/SubaruProject/SXDS/index.html}).
\end{abstract}



\section{INTRODUCTION}\label{sec:intro} 
Understanding the formation and evolution processes of the individual
galaxy and the growth of the large scale structures (LSSs) in the universe
is one of the major goals in extragalactic astronomy today. 
In the scheme of a typical $\Lambda$-CDM model, the growth of structures is 
governed by the gravitational growth of initial fluctuations of dark matter.
The baryonic material cools in these dark matter structures and 
grows through hierarchical clustering to galaxies and clusters of galaxies.
The subsequent evolution of galaxies is closely connected to their environments,
which, of course, relate to the LSSs, where those galaxies are located.
The ultimate goal of extragalactic surveys is to trace these evolutionary processes
by well-defined statistical galaxy samples.

Optical imaging is arguably the cornerstone of any extragalactic
survey, since it provides identifications and positions of celestial
objects for follow-up spectroscopy. In recent years, there have been a
number of deep imaging survey projects which devote significant amounts
of telescope time, such as Great Observatories Origins Deep Survey
(GOODS: Giavalisco et al. 2004) and Cosmic Evolution Survey (COSMOS: HST treasury
project: Feldmann et al. 2006; Scoville et al. 2007).
These surveys provide multi-waveband galaxy samples at faint magnitude.
Then, the photometric redshift techniques (Furusawa et al. 2000;
Bolzonella et al. 2000; Feldmann et al. 2006) are
frequently used to pre-select candidates for spectroscopy. 

Although a great deal of information can be obtained from the optical
data alone, the value of the data set grows significantly as data at
other wavelengths from other facilities are added. We therefore elected
to use a multi-wavelength approach for the Subaru/XMM-Newton Deep Survey
(SXDS; Sekiguchi et al. 2007, in preparation) from the very start,
ensuring that our chosen field would be accessible and suitable for
observations at all wavelengths.
An equatorial field, centered at R.A. =
$02^{h}18^{m}00^{s}$, decl. = $-05^{\circ}00'00''$ is chosen to tie up
with the deep X-ray observations with XMM-Newton observatory 
(Ueda et al. 2007), and also due to the accessibility by all major
observatories, both existing and planned.
The major multi-wavelength observation programs on the SXDS field
(hereafter SXDF) include deep radio imaging with the VLA (Simpson et
al. 2006), sub-mm mapping with SCUBA (Coppin et al. 2006; Mortier et al.,
2005), mid-infrared observations with Spitzer Space Telescope (Lonsdale
et al. 2003), deep near-infrared imaging with the UKIRT/WFCAM (Foucaud
et al. 2006; Dye et al. 2006; Lawrence et al. 2007), and the X-ray
observations with XMM-Newton observatory. Importantly, the survey field
of an infrared ultra deep survey (UDS; Foucaud et al. 2006) covering 0.77
square degrees as part of the UKIDSS project (Lawrence et al. 2007) is
centered on the SXDF.
It is expected that our extensive multi-wavelength data set will provide photometric
redshifts accurate to $\Delta z\lesssim 0.1$ over a wide range of
redshift, as well as detailed spectral energy distributions for the vast
majority of the objects in the field.

The SXDS has been undertaken as a part of the Subaru Telescope
``Observatory Projects'', in which a large amount of observing times are
devoted to carry out intensive survey programs by combining observing
times rewarded to builders and observatory staff of the Subaru
Telescope.
Note that Subaru Deep Field (SDF; Kashikawa et al. 2004) is another
observatory project, which targets on a field different from the SXDF. 
The data set of the SDF is slightly deeper than the SXDF, but concentrates
on one-fifth narrower field than the SXDF. 
Thus, these surveys complement each other and they can be used for a
wide variety of studies. 

In this paper, we describe the observation details and data reduction
and analysis of our optical imaging survey with the prime-focus camera on Subaru Telescope, 
as well as the resultant data products.
A detailed description of the survey strategies, scientific objectives,
and multi-wavelength survey plans is given in a companion paper
(Sekiguchi et al., 2007 - Paper I, in preparation).
In Section~2, we describe optical imaging observations by
Subaru Telescope, and in Section~3 we explain the data reduction procedure
employed.  We present creation of multi-waveband photometric catalogs 
including object detection and aperture photometry in Section~4, and 
calibrations of photometry and astrometry in Section~5.
In Section~6, we discuss characteristics of the catalogs such as
limiting magnitude, detection completeness, and number counts of
galaxies.  The summary is given in Section~7.  Throughout this paper,
all the magnitudes are expressed in the AB system unless otherwise mentioned.

\section{OBSERVATIONS AND DATA}\label{sec:data} 
\begin{figure}
\epsscale{1}
\plotone{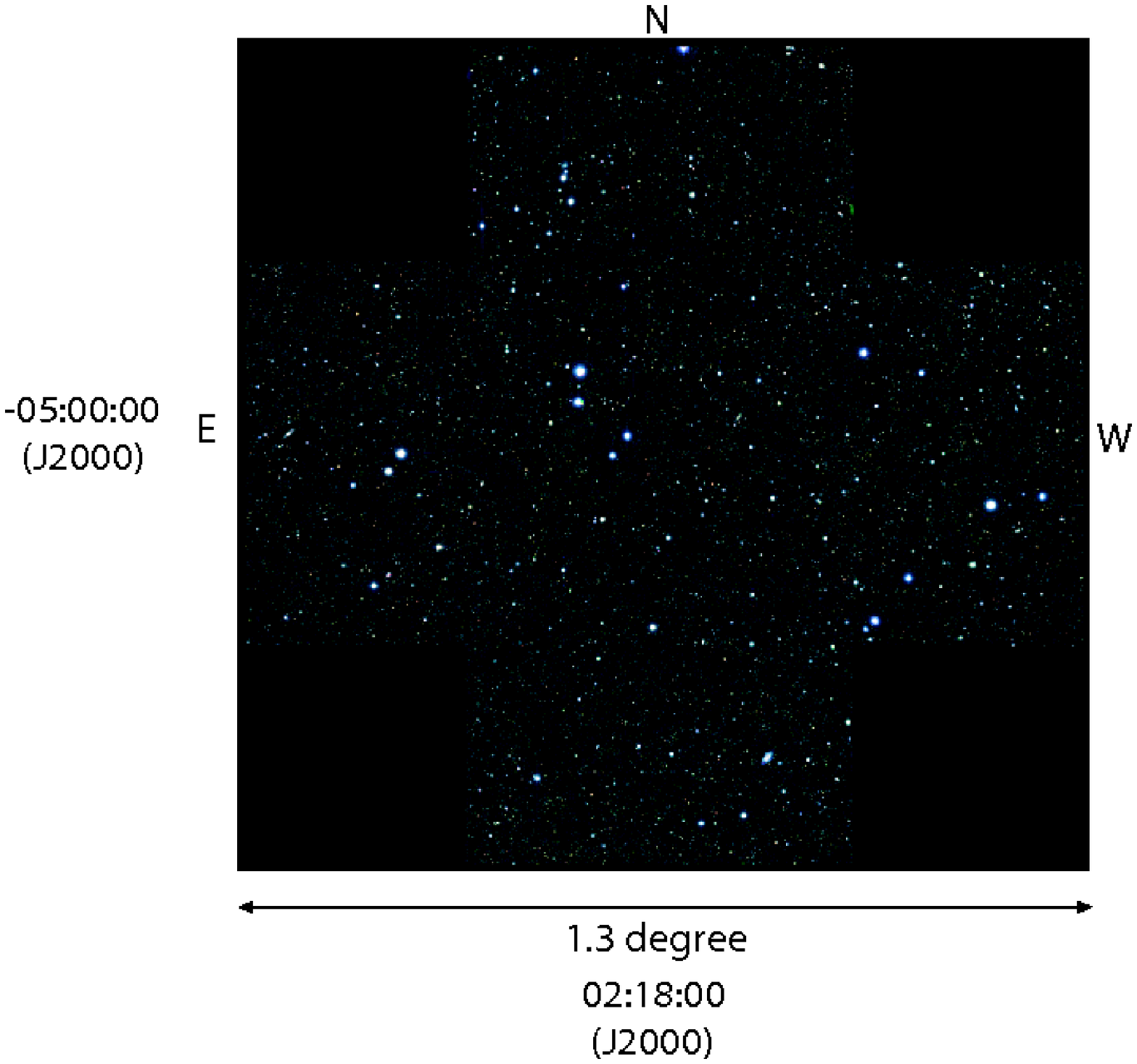}
\caption{Three color ($B$-, $R_c$- and $i'$-band) composite optical image of the Subaru/XMM-Newton Deep survey Field (SXDF). The area consists of five Suprime-Cam pointings, corresponding to a 1.22 square degree field.}
\label{fig:pseudo-img}
\end{figure}

The optical imaging observations of the SXDF are carried out using 
the prime-focus camera (Suprime-Cam; Miyazaki et al. 2002)
on Subaru Telescope in the period from September 2002 to September 2005.
Suprime-Cam is a $5\times2$ mosaic of ten MIT/LL buttable 2k$\times$4k CCD camera.
It has a $34'\times 27'$ field of view with a projected pixel scale of
0.202 arcsec.
The layout of the pointings is arranged as a cross shape so that each of
the North-South and the East-West directions has an extent of $\sim 1.3$
deg (Figure~\ref{fig:pseudo-img}).
This corresponds to a field span of a transverse dimension of $\sim 75$
Mpc at $z\sim 1$ and $\sim 145$ Mpc at $z\sim 3$ in the comoving scale
for ($h, \Omega_M, \Omega_\Lambda)=(0.7, 0.3, 0.7$).
The coverage of the comoving volume is $\sim3.2\times 10^{7}$ Mpc$^{3}$ for $z=0-3$.
Table~\ref{tab:coords} lists field center coordinates of the five pointings.
Hereafter, the five pointings are referred to as $C$ (Center), $N$ (North),
$S$ (South), $E$ (East), and $W$ (West), with respect to their relative positions
on the sky.

\begin{figure}
\epsscale{0.8}
\plotone{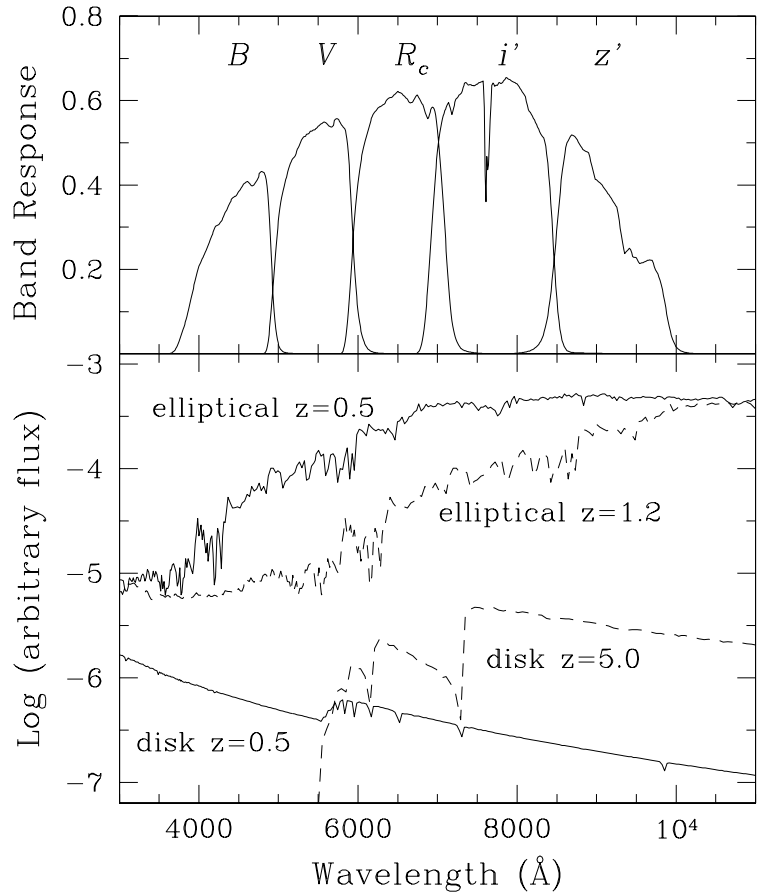}
\caption{Band response curves of the SXDS Suprime-Cam imaging observations (upper panel) and the typical spectra of an early-type galaxy (labeled as ``elliptical'') and a late-type galaxy (labeled as ``disk'') at the different redshifts (lower panel).}
\label{fig:band}
\end{figure}

Observations of the SXDF are performed using five broad-band filters,
$B$, $V$, $R_c$, $i'$, and $z'$ to cover the entire wavelength range
observable with Suprime-Cam.
The transmission curves of the five filters employed in the SXDS, convolved with
the CCD sensitivities, reflectivity of the primary mirror, transmissions 
of corrector lenses, and atmospheric transmission, are given in
Figure~\ref{fig:band}.
Also, given in Figure~\ref{fig:band} are the redshifted spectral energy
distributions (SEDs) of an early-type galaxy at $z=0.5$ and $1.2$ and a
late-type galaxy at $z=0.5$ and $5.0$ for references.
With photometric data sampled in the 5 bands, we can investigate color
properties of red passively evolving galaxies to $z\sim 1$ (see Kodama
et al. 2004; Yamada et al. 2005) and young star-forming galaxies at high
redshift such as lyman break galaxies (LBG) (Ouchi et al. 2004, 2005).
The observations used in this paper span observing runs over a period of 3 years.
A total of 160 hours are allocated to the SXDF imaging
observations as a part of Subaru Telescope ``Observatory Projects'' (see
also Kashikawa et al. 2004) combined with observatory staff times. 
In addition to these times, we use 14 hours of the observing times 
offered by the Supernova Cosmology Project (Doi et al. 2003; Lidman et
al. 2005), which add the total exposure time of the center and the west
pointings in the $i'$ band ({\it $i'$-C} and {\it $i'$-W}). Moreover, in
the course of another survey program for Ly$\alpha$ emitting galaxies
(Ouchi et al. 2005, 2007), 3 hours are devoted to the $V$-band
imaging. In total, 133 hours are used for on-source exposures for the
final data set described in this paper. The complete log of the
observations is given in Table~\ref{tab:log}.

\begin{figure}
\epsscale{1}
\plotone{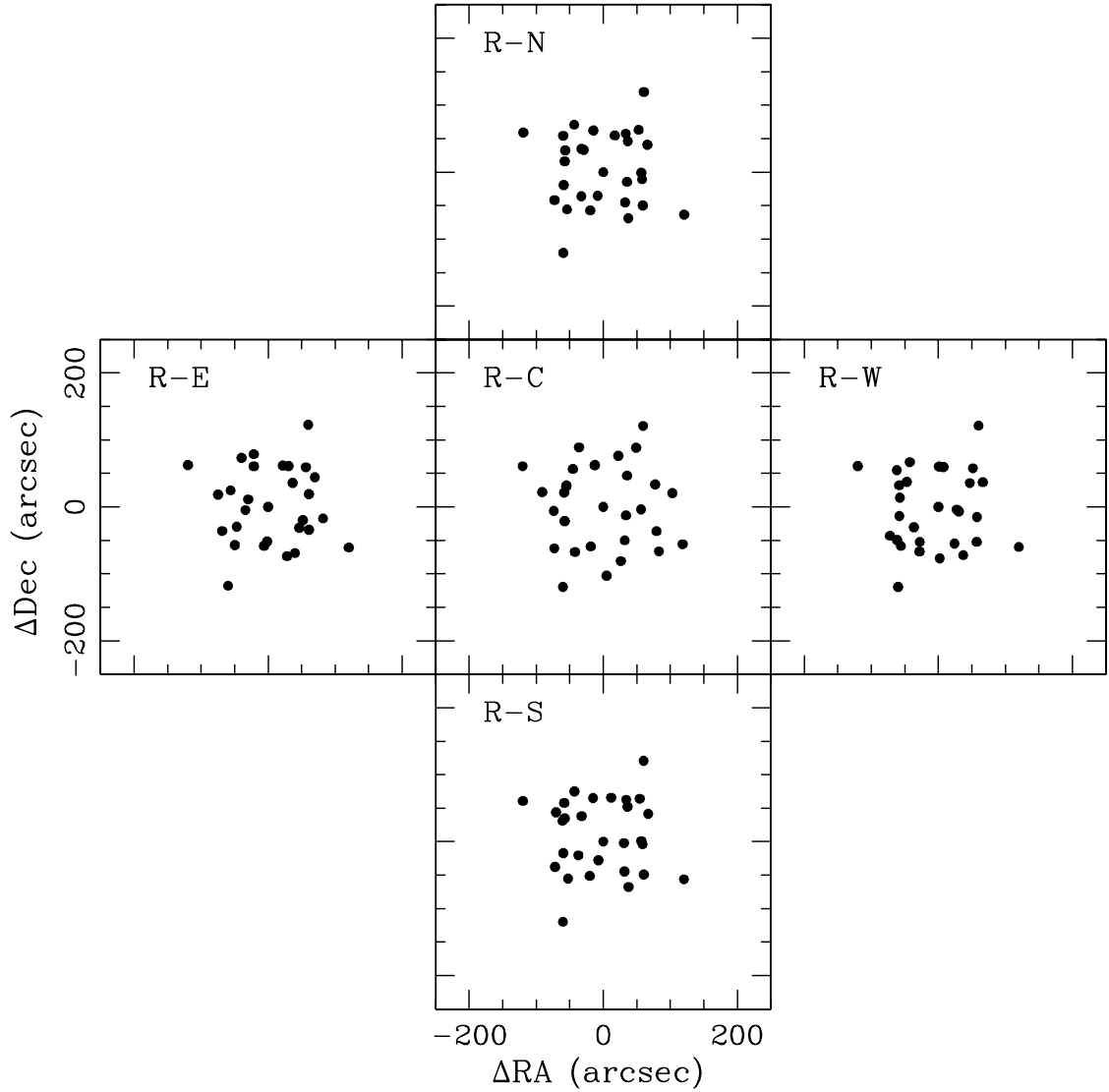}
\caption{Dithering patterns for the $R_c$-band observations for five
 pointings. The center $(0,0)$ position of each pointing corresponds the
 coordinates given in Table~\ref{tab:coords}.} 
\label{fig:dither}
\end{figure}

The coordinates of the five pointings are carefully chosen so that 
the resultant images uniformly cover the entire SXDF, though 
we analyze each of the pointings separately in the present study.
Also, to eliminate bad pixels, gaps between individual CCD chips and cosmic ray events,
circular dithering pattern is used for each pointing.
This dithering pattern employs a circular motion with radii of 60
to 120 arcsec around the center position of each pointing, combined with 
slight offsets to the center positions of the circular motion. 
As an example, the dithering pattern for the $R_c$ band 
is plotted for each pointing in Figure~\ref{fig:dither}.

\section{DATA REDUCTION}\label{sec:reduc} 
The SXDS images obtained are processed using an in-house pipeline
software package (See Ouchi 2003). We use it in almost the same way that
successfully used for the SDF (Kashikawa et al. 2004).
This pipeline is based on a software package developed for
Suprime-Cam data reduction (Yagi et al. 2002) and several IRAF
tasks (geomap, geotran, etc.) are used in the process of a geometrical
transformation and alignment of final stacked images.
We use a standard mosaic-CCD data reduction procedure, which is briefly
summarized below. 
 
First, we apply a bias correction to the raw data.
The median count of the overscan region is computed for each line, 
which represents a typical bias level at that line, 
and this median count is subtracted from the counts in all the pixels in that line.
The overscan regions are trimmed after the bias subtraction.  
We assign a flag number (-32768) to the saturated or detected bad pixels, and 
these pixels are ignored in the following reduction processes. 
Since pixels in the outside edges of each frame are likely to be
affected by noises which can not be corrected by the following
flatfielding process, they are also masked with the flag number.

Flat frames are constructed from the normalized object frames taken
during the same observing run. We use more than 150 object frames to
create flat frames.  
If the number of object frames is less than 150 frames, object frames
taken in different observing runs within 3 months or twilight flat
frames taken in the same run are combined 
to increase the signal-to-noise ratios (S/Ns) of the flat frames for
that period. 
The median value of the each pixel is used to create the flat frames.
Pixels which could be affected by bright objects and vignetting by the
auto-guider probe are flagged out before processing the flat frames.
Then, object frames are divided by the flat frames.
The peak-to-peak ratios of background levels over the entire
flat fielded frames are less than $2\%$. 
We do not expect any significant systematic errors by the
flatfielding that affect accuracies of the following analyses.

After the flatfielding process is done, the distortion correction is
applied to each frame using the 5-th order polynomial transformation
formula derived from Miyazaki et al. (2002).
The changes in shapes and sizes of objects in the frames due to the
distortion are negligible compared with the size of PSFs.  
Each pixel is transformed so that the surface brightness of the pixel 
is conserved, since the relative flux in each pixel has been already
corrected by the flatfielding process.
Positional displacement of objects along the perpendicular orientation 
due to the atmospheric effect is also corrected at the same time.

The sky background is subtracted from each distortion-free frame, 
We divide an image into temporary meshes with $64\times 64$ pixels,
corresponding to 12.9 arcsec in one direction. 
The mode of counts in each mesh 
is adopted as the sky background levels after smoothed by a median
filter with a  $3\times 3$ kernel.
This global sky subtraction should not affect photometric accuracies 
of compact objects such as galactic stars and faint galaxies.
However, caution must be taken when we apply these values to 
extended objects such as nearby galaxies with low surface brightness. 

The stacking of the images is performed for each of the pointings
separately in the following way.
First, we choose 100 to 200 unsaturated stellar objects with a peak flux
of $>500$ADU in each frame distributing evenly over the entire frame.
Second, we determine a reference frame of all the reduced frames.
For the other frames, relative positions and rotation ($\Delta x,
\Delta y, \Delta \theta$) and flux ratios to the reference frame are 
determined based on a least square method using these stellar objects
commonly detected in the frames. 
Third, for all the frames, the positional shifts and rotations 
are corrected.  
Here, we do not use frames which have the flux ratios of less than 
$30\%$ of that of the reference frame.
Finally, all the frames are co-added by calculating flux-weighted
average values in each pixel with a 3-$\sigma$ clipping process. 
The clipping process effectively removes cosmic rays and bad or saturated 
pixels, and satellite trails etc., and should not affect the resulting 
total fluxes of objects because the difference in counts among 
frames with various seeing sizes is mostly within the 3-$\sigma$
threshold.

After the co-adding of each band images, all the five-band ($B, V, R_c,
i', z'$) images are registered into identical positional
coordinates as we intend to execute multi-waveband aperture photometry.
The positional registration is performed by the geometrical
transformation by 3-rd order polynomials fitting using the positions of stellar 
objects common to the co-added stacked images. 
The r.m.s. deviation of residuals of transformed positions with respect to the 
reference positions in the $R_c$-band images is approximately 0.1-0.15 arcsec.

For each pointing, the geometrically matched images are smoothed 
with Gaussian filters iteratively so that the PSF sizes of all the
images become the same as that in the worst image used in this study.
The PSF sizes of the resultant images are $0.80, 0.84, 0.82, 0.82$, and
$0.82$ arcseconds in {\it SXDS-C, SXDS-N, SXDS-S, SXDS-E}, and {\it SXDS-W},
respectively.

Finally, regions with low S/Ns or those affected by saturation or
overflow of electrons are carefully checked and assigned as flagged
regions which are indicated in photometric catalogs. 
The resultant PSF sizes and total usable areas of each pointing is
summarized in Table~\ref{tab:stacked}.

\section{MULTI-WAVEBAND PHOTOMETRIC CATALOGS} 
We construct multi-waveband photometric catalogs of the objects
identified in the SXDF.

\subsection{Object Detection}\label{sec:object_detection}
Object detection and photometry are performed using SExtractor version
2.3.2 (Bertin \& Arnouts 1996). 
Before performing the object detection and photometry, the outskirts of
the stacked images are trimmed in order to eliminate any failure such as
inappropriate estimation of sky levels and memory overflow during extraction of
objects by SExtractor due to a high noise level.  
After trimming the outskirts, each individual image has similar size
of approximately $10,800\times 8,100$ pixels.  
We use the detection criteria that at least five pixels which have
values of $> 2\sigma$ of sky background r.m.s. noise must be connected
in order to be included in our object list.  Positions of detected
objects are translated to sky coordinates based on the astrometric
calibration which will be discussed in Section~\ref{sec:astr_calib}.

\subsection{Aperture Photometry}\label{sec:apphot}
After the object detection is completed in one of the five bands, which
we call `detection band', multi-waveband aperture photometry is
carried out with SExtractor for the other four bands using the same 
apertures at the same positions as those in the detection band.  
We perform aperture photometry with fixed circular apertures ({\tt MAG\_APER}) of
$2$ and $3$ arcseconds in diameter, and variable elliptical apertures
({\tt MAG\_AUTO} and {\tt MAG\_BEST}) utilizing the Kron's first moment scheme with
apertures' radii of $2.5\, r_\mathrm{Kron}$ (Kron 1980). 
Measured fluxes of objects are converted into magnitude using the
photometric zeropoints, which will be described in Section~\ref{sec:photo_calib}.
The extracted magnitude is considered to be the asymptotic
total magnitude of galaxies, which should sample at least $94\%$ 
of flux of a galaxy (Bertin \& Arnouts 1996). 
Other useful parameters for objects including shape, size, and
stellarity and so on, are also extracted by SExtractor at the same time.
All the object parameters which are listed in the catalogs are
summarized in Table~\ref{tab:catalog}.

In this procedure, $B$-, $V$-, $R_c$-, $i'$-, and $z'$-selected multi-waveband catalogs
are created for each pointing separately, which means that each object 
has measurements in the five bands ($B, V, R_c, i', z'$) in each
catalog.  Thus 25 catalogs (= 5 pointings times 5 detection bands) in
total are created.  

\subsection{Flagged Areas}\label{sec:flags}
Although we have flagged out bad pixels from the image frame in the
reduction procedure, there still remain (1) areas strongly affected mainly by
overflow of electrons or envelopes of extended objects, and (2) areas
weakly affected by large halos from very bright stars or noisy regions. 
In our catalogs, the objects which are located in these areas are
flagged 1 and 2, respectively. 
Otherwise, the objects located in the clean area have a flag value of 0.

The basic parameters used in the object detection and aperture
photometry are summarized in Table~\ref{tab:sexparam}.
The complete configuration files for SExtractor can be found at {\tt
http://www.naoj.org/Science/SubaruProject/SXDS/index.html}. 
Note that we adopt default parameters of SExtractor for the source
extraction unless we do not find any problem with the output 
values.  The parameters adopted are suitable for extraction of moderate
to faint compact galaxies, and not heavily optimized to extraction of 
extended objects or very faint objects. Thus the reliability of the current
catalogs for those objects should be regarded with some caution.

\section{CALIBRATIONS}\label{sec:calib} 
\subsection {Photometric Calibration}\label{sec:photo_calib}
Photometric zeropoints in all the five bands are determined in the following way.
First, we compare our observation with the earlier calibrated
observations for the same overlapping field, which were made by the
Suprime-Cam instrument development team (Ouchi et al. 2001, 2004).  
The southern half of our center pointing ({\it SXDS-C}) is 
overlapped with the northern half of the image obtained by Ouchi et al. (2001).
They determined the photometric zeropoints in the Suprime-Cam band
system based on observations of photometric standard stars SA92 and SA95
for $B, V$ and $R_c$ bands and a spectrophotometric standard star SA95-42 for
$i'$ and $z'$ bands.

We estimate the photometric zeropoints for our center pointing ({\it SXDS-C}) 
by comparing magnitudes ({\tt MAG\_AUTO}) of $\sim 200$ objects which
are commonly detected in the two images and have FWHMs of smaller than
1.2 arcsec with a magnitude range of 20.0 to 24.0.
Then, we extend these zeropoints to the surrounding four pointings by
using objects in the overlapping areas in the same manner as for the
determination of the zeropoint for the {\it SXDS-C}.
In our catalogs, we list the instrumental magnitude obtained 
in the Suprime-Cam band system. Hence, no transformation of magnitude
considering color terms of band responses are conducted.
The color terms between the Suprime-Cam band system and the standard
system are basically small -- for the reddest stars, of order 0.1 in the
$B,V,i',z'$ bands, and slightly large ($\sim0.3$) in the $R_c$ band.

\begin{figure}
\epsscale{1}
\plotone{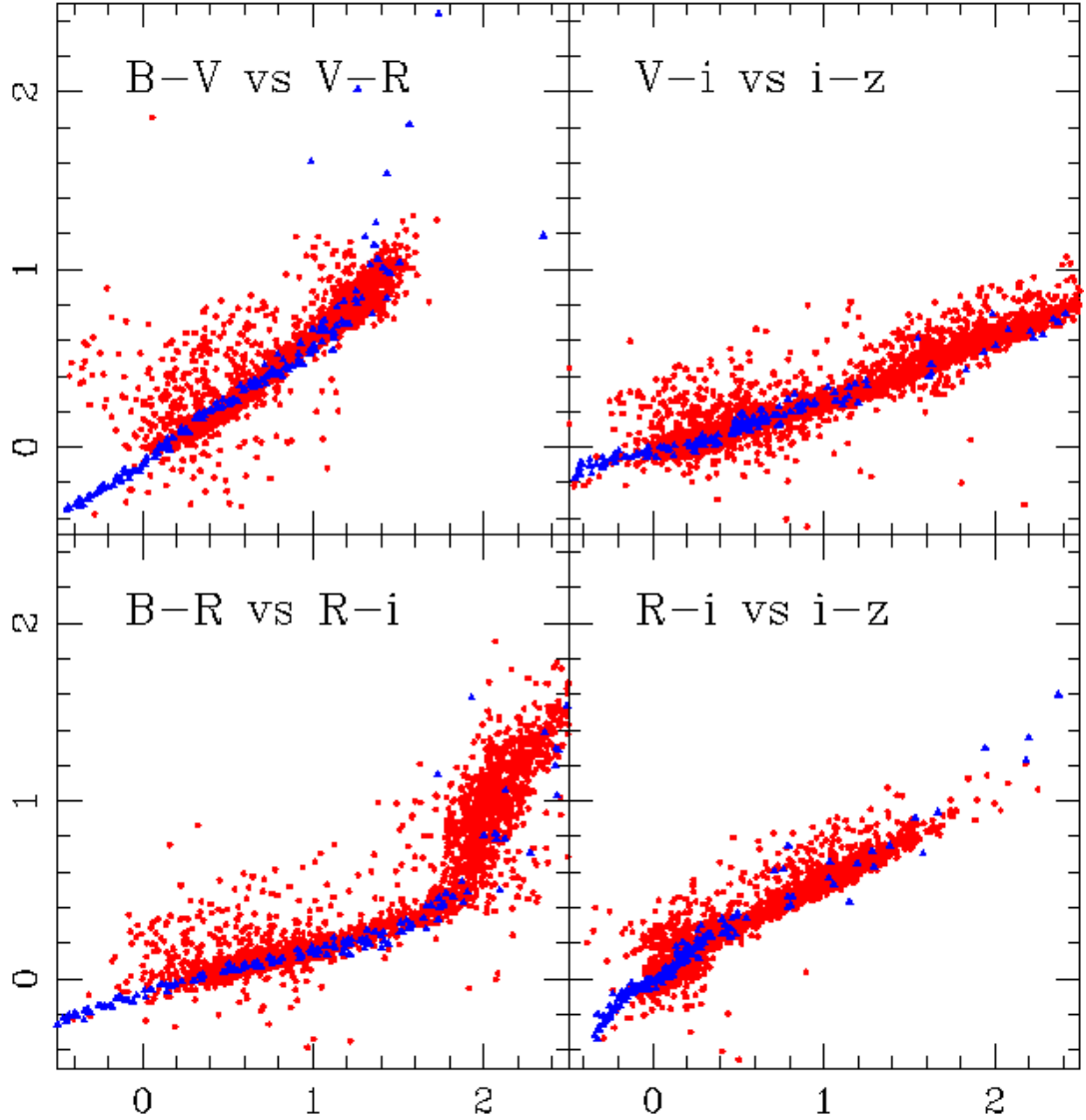}
\caption{Color-color plots for the stellar objects in the SXDF (red dots) and Gunn-Stryker stars (blue triangles). The SXDF colors agree with the colors of Gunn-Stryker stars to the accuracy of 0.03 mag or better.}
\label{fig:ccplot_gs}
\end{figure}

To test internal consistency of our zeropoint calibrations, 
we examine colors of stellar objects in our photometric 
catalogs and those computed by convolving typical 
stellar SEDs covering the complete ranges of spectral types (Gunn \&
Stryker 1983) with Suprime-Cam response curves.  
We find that it is necessary to apply corrections to the zeropoints 
of $\Delta V=+0.03$ and  $\Delta z'=+0.03$ for all the five pointings, 
and $\Delta i'=+0.05$ and $+0.02$ for the South and East pointings 
({\it $i'$-S} and {\it $i'$-E}), respectively. 
After these corrections are applied, the colors of the stellar objects
in our catalogs agree with those computed for the SEDs from 
Gunn \& Stryker (1983) within $0.03$ mag in all the five bands 
(Figure~\ref{fig:ccplot_gs}).
The systematic errors adjusted here may be due to relatively small
number of stellar objects used in the determination of the photometric
zeropoints.

To ensure the accuracy of our zeropoints, we conduct CCD photometric
observations of a part of the SXDF in $B, R_c, i'$ and $z'$ bands
with the UH88-inch telescope at Mauna Kea on October 9 and 10, 2002.
We compare magnitudes of objects in $B, R_c, i'$ and $z'$ bands obtained
 by the UH88-inch observations with our catalog values. We find that our
 catalog values agree within 0.05 mag r.m.s. for $B$ and $R_c$ bands
 and within 0.1 mag r.m.s. for $i'$ and $z'$ bands.
The relatively large difference seen in $i'$ and $z'$ bands are probably due
to the difference in the band responses between the UH88-inch system and
the Suprime-Cam system, and low S/Ns of shallow SXDS images used for 
comparison with the UH88-inch data.
To confirm accuracy of the photometric zeropoints in the 
$i'$ and $z'$ bands in our catalogs, follow-up observations are made for
the two bands with the 1.0-m telescope at the United States Naval
Observatory (USNO). 
The observations at the USNO were performed on the photometric night of
December 3, 2004 in the course of photometric calibration for a high-$z$
supernovae search (Yasuda et al. 2007, in preparation).
We find that a systematic difference between zeropoints in our catalogs
and those determined by the USNO observation is as small as $0.03$ mag
r.m.s.

Thus we conclude that the uncertainties of calibrated photometric
zeropoints of the SXDS images are $0.03-0.05$ mag r.m.s. The adopted
photometric zeropoints are listed in Table~\ref{tab:log}.

\subsection {Astrometric Calibration}\label{sec:astr_calib}
Astrometric calibration is performed using $\sim$200 stars 
with magnitude of $J_\mathrm{vega}<16.5$ derived from 2-MASS point
source catalogs (PSC; Cutri et al. 2003), in each of the five stacked
$R_c$-band images.
Since images in the other bands have been geometrically transformed, 
the positions of objects detected on the images should coincide with 
those found in the $R_c$-band images (see Section~\ref{sec:reduc}). 
The world coordinates for the images are calculated for the $R_c$-band
images and written in the FITS headers of all the stacked images 
with the IRAF tasks CCMAP and CCSETWCS.
The r.m.s. uncertainties of the determined coordinates across the FOV
with respect to the PSC positions are on the order of 0.2 arcsec in both
RA and Dec in each pointing (Table~\ref{tab:astmt}). 
The positional error of the PSC is reported as $\sim 0.07$ arcsec.

\begin{figure}
\epsscale{1}
\plotone{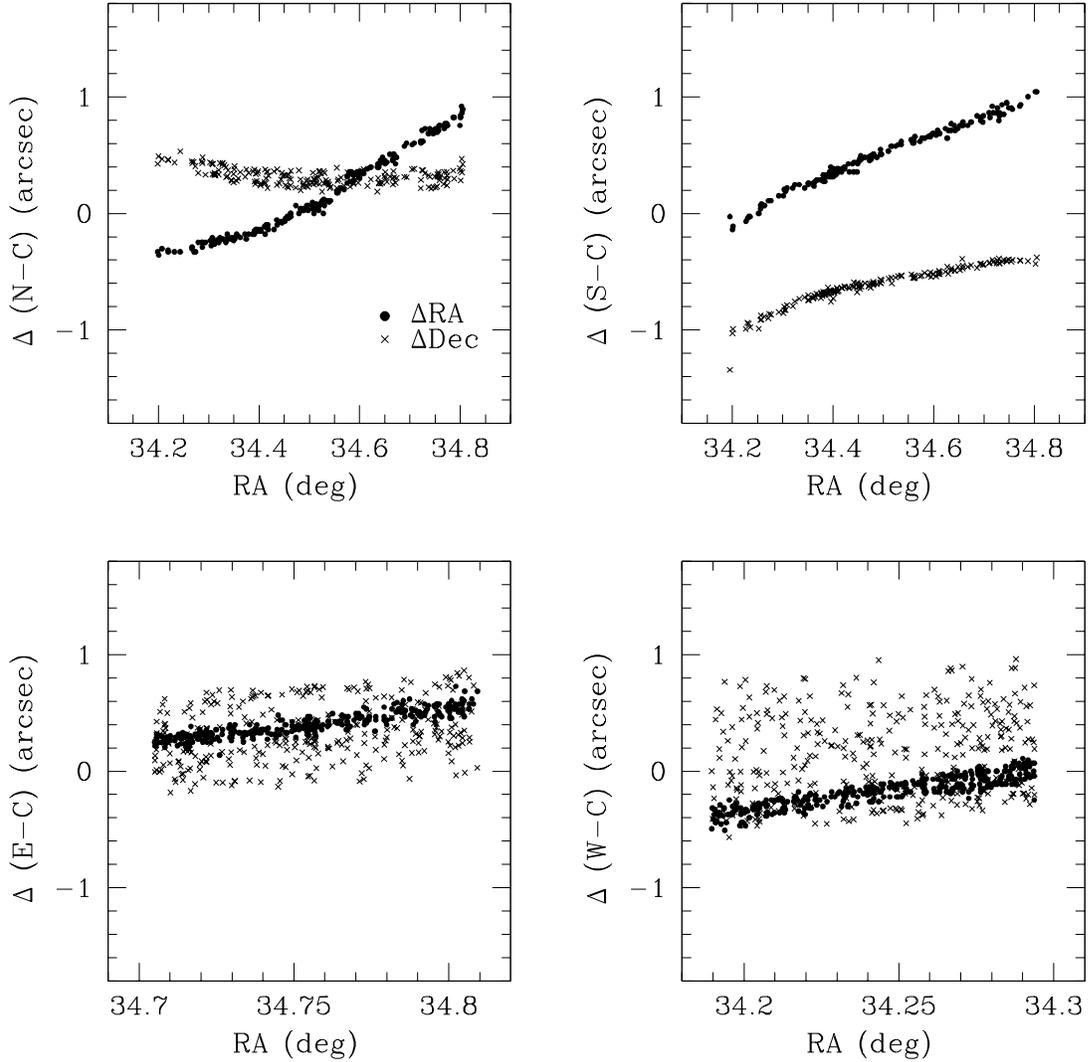}
\caption{Residuals in the RA (filled circles) and in the Dec (crosses)
 of the World-Coordinate values in the RA direction in 4 overlapping
 areas between the center pointing ({\it SXDS-C}) and the north ({\it
 SXDS-N}: upper left panel), {\it SXDS-C} and the south ({\it SXDS-S}:
 upper right panel), {\it SXDS-C} and the east ({\it SXDS-E}: lower left
 panel), and {\it SXDS-C} and the west ({\it SXDS-W}: lower right
 panel). Only the objects with the FWHM$<1.2$ arcsec are plotted.}
\label{fig:astmt_diff1}
\end{figure}

\begin{figure}
\epsscale{1}
\plotone{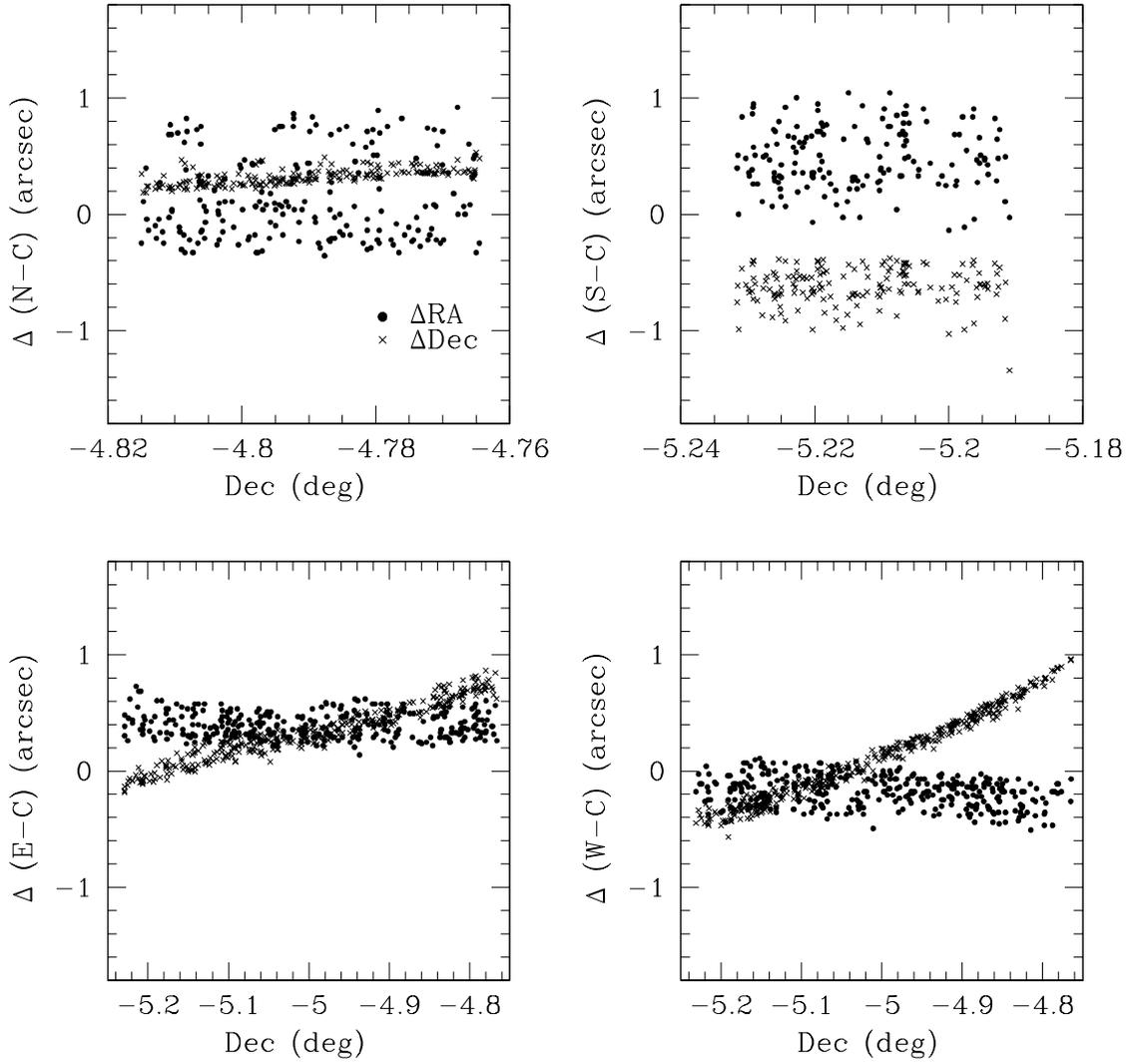}
\caption{Same as Figure~\ref{fig:astmt_diff1}, but the residuals in the Dec direction.}
\label{fig:astmt_diff2}
\end{figure}

We examine possible systematic differences in the calculated world
coordinates between different pointings by using objects in the
overlapping areas of two pointings.  Figure~\ref{fig:astmt_diff1} and
Figure~\ref{fig:astmt_diff2} show the differences in RA and Dec of
point-like objects with FWHMs $<1.2$ arcsec in the magnitude 
range $R=20.5-24.0$ detected both in the {\it SXDS-C} image and other
surrounding images. 
From the figures, the world coordinates assigned to each of the five
pointing images are in good agreement with one another with accuracy of
about 1 arcsec at the edges of the images.
The differences of the world coordinates between two pointings
studied here are thought to be for the worst cases, 
since they are derived based on only the objects in the overlapping 
areas which are located at the edge of each image.
Slopes and offsets in the residuals are seen in each panel of the figures.
These disagreements in the positions of objects between two overlapping 
images come from possible tilts, offsets and/or difference in
geometrical scale of the world coordinate systems between the two images.
Residuals of distortion in the images and the atmospheric effect, which
have not been completely removed in the reduction process, may cause 
the errors in determination of the world coordinate systems. 
Nevertheless, the SXDS catalogs have a good enough positional accuracy
to perform follow-up spectroscopy. 

The astrometrically and photometrically calibrated images of the SXDS
in five bands ( $B, V, R_c, i', z'$) described in this section are
released via SXDS Web Page ({\tt http://www.naoj.org/Science/SubaruProject/SXDS/}). 
The data set described in this paper is based on the data release 1 (DR1). 
A complete history of the data release is summarized in the web site.
The raw image data are also available to the public via Subaru Telescope ARchives System
(STARS: {\tt https://stars.naoj.org}).


\section{PROPERTIES OF THE CATALOGS}\label{sec:properties_catalog}  

\subsection{Galactic Extinction}\label{sec:galactic_extinction}
All the magnitudes listed in the catalogs are not corrected 
for the Galactic extinction.  The magnitude attenuation in each band 
estimated for the central position ($02^{h}18^{m}00^{s},
-05^{\circ}00'00''$; J2000) based on Schlegel et al. (1998), assuming an
extinction curve with $R_V=A_V/E(B-V)=3.1$ is as follows: $A_B=0.091,
A_V=0.070, A_{R_c}=0.056, A_{i'}=0.044$, and $A_{z'}=0.031$. 
Table~\ref{tab:gal_ext} summaries the magnitude attenuation in each band for all the pointings.
The Galactic extinction above must be taken into account when studying
extragalactic objects. 

\subsection{Limiting Magnitudes}\label{sec:limmag}
We examine 3-$\sigma$ limiting magnitudes measured for the
$2''$-diameter apertures in all five bands ($m_{lim}$ column in
Table~\ref{tab:log}) in the following manner. First, counts in unit of
ADU fallen into the $2''$-diameter apertures are measured at approximately
10,000 positions randomly selected and spreading over the entire image.
Next, a histogram of the counts is produced and only the faint side of
the histogram is fit by the Gaussian function, as the bright-side tail
is composed of not only the sky background but also photons from the objects. 
The $\sigma$ of the best-fit Gaussian is regarded as the
1-$\sigma$ sky fluctuation of the image for the $2''$-diameter apertures. 
We find that limiting magnitude in each band is $B=28.4, V=27.8,
R_c=27.7, i'=27.7$ and $z'=26.6$ ($AB,  3\,\sigma, \phi = 2''$)
in the deepest images of the five pointings.

\subsection{S/N Distribution Map}\label{sec:snmap}
To examine the homogeneity of S/Ns of the SXDS data across the
entire field, we investigate the S/N distribution map in the $R_c$-band
images. 
First, we calculate the sky fluctuation per $2''$-diameter aperture 
within a mesh with a size of $350\times 350$ pixels at each position in
the same manner as in estimation of the limiting magnitudes.
The mesh is shifted with a step of 175 pixels to cover the entire field
of view.
Then, the fluctuations for the 2''-diameter aperture are converted to
S/Ns for objects with an $R_c$ magnitude of 27.5 at each position.

\begin{figure}
\includegraphics[scale=0.3]{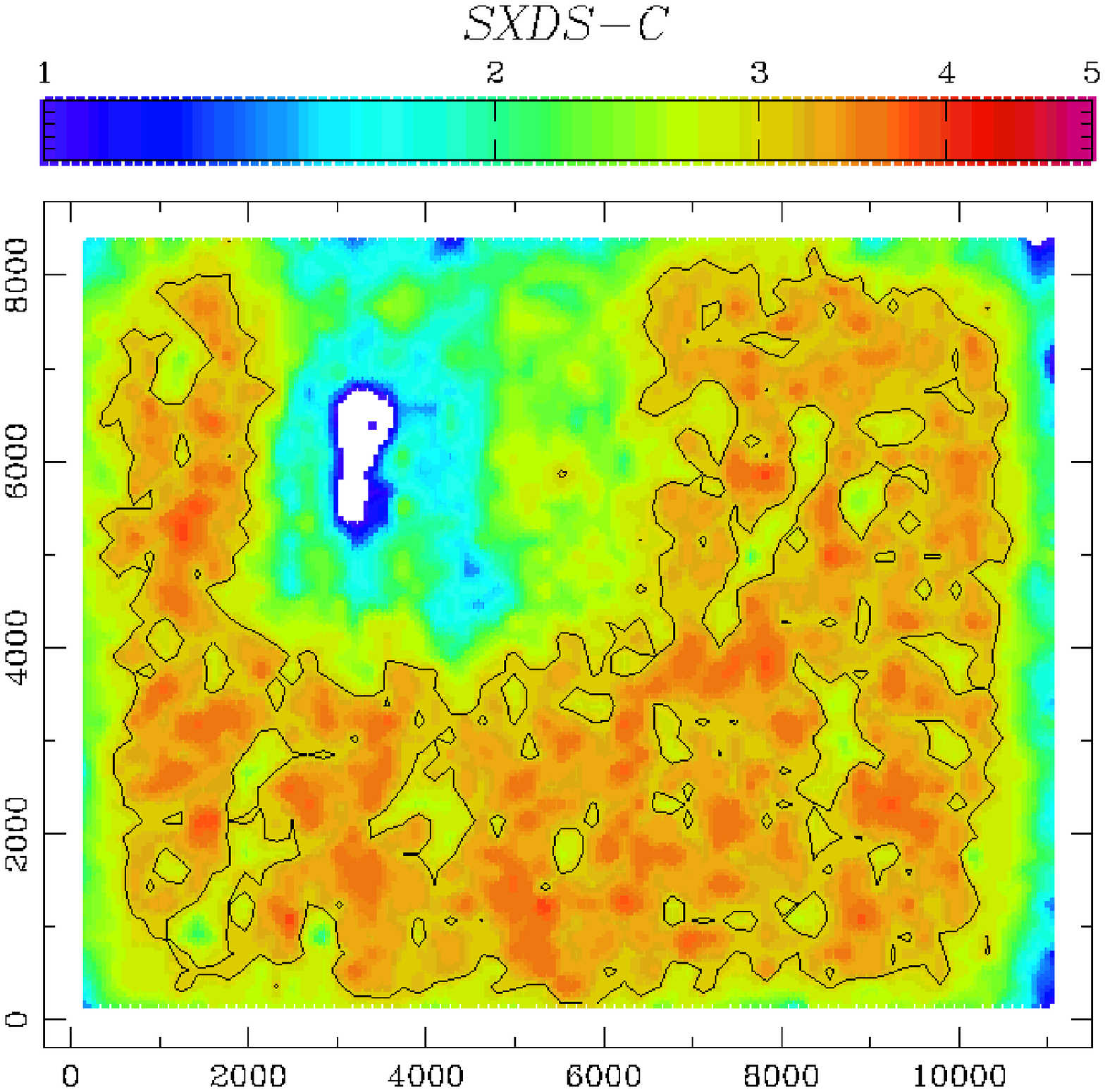}
\includegraphics[scale=0.3]{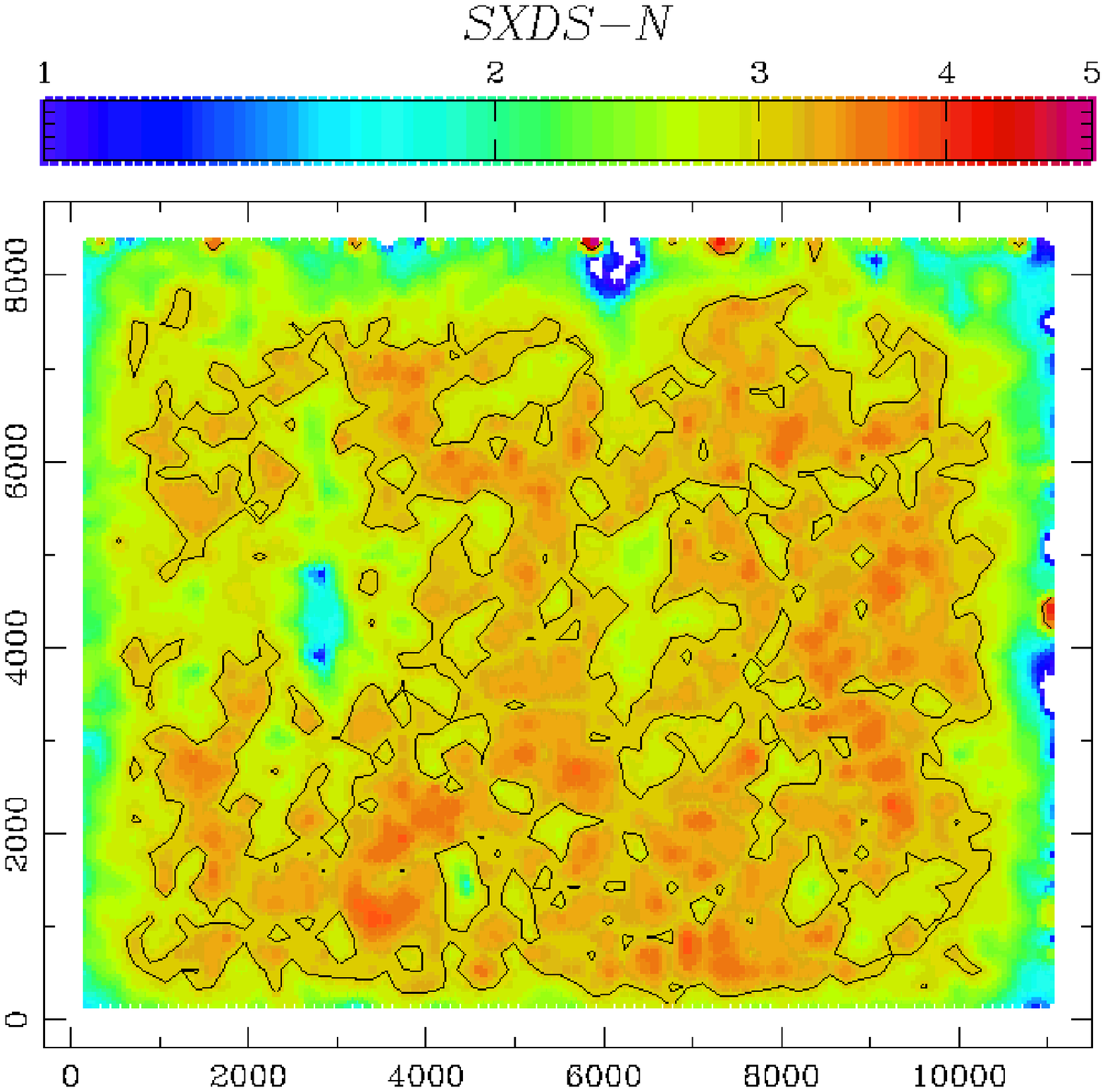}
\includegraphics[scale=0.3]{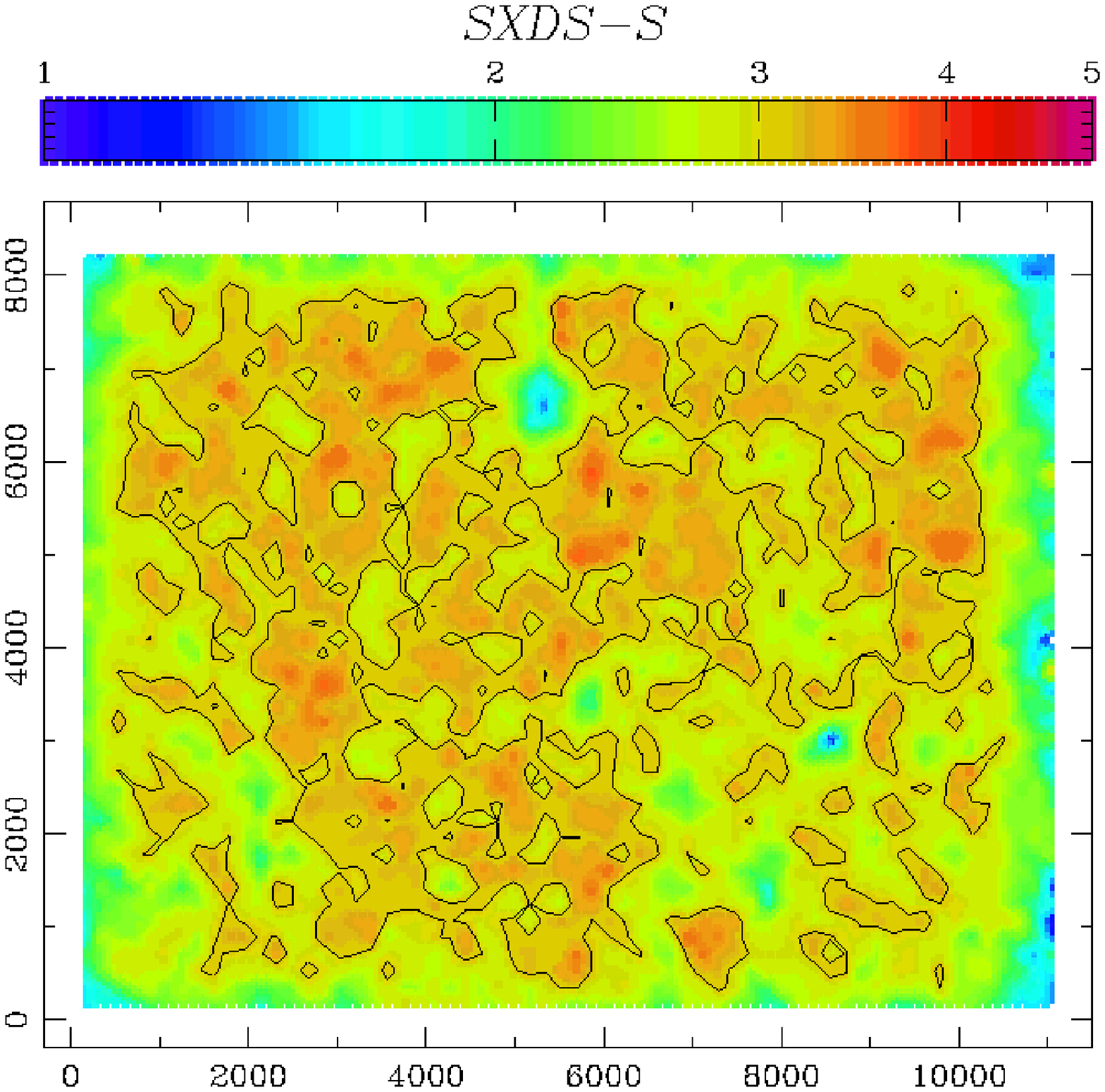}
\includegraphics[scale=0.3]{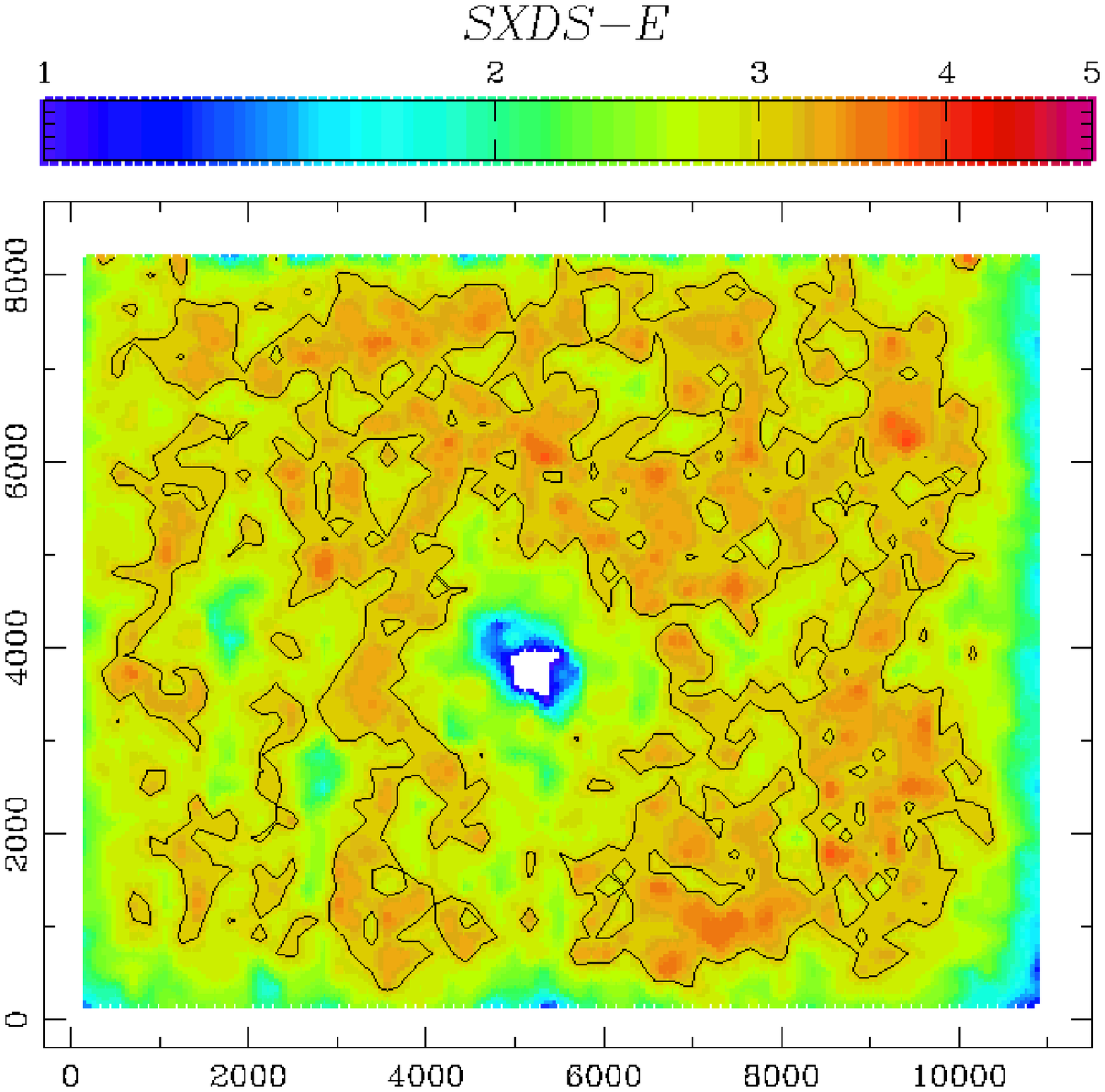}
\includegraphics[scale=0.3]{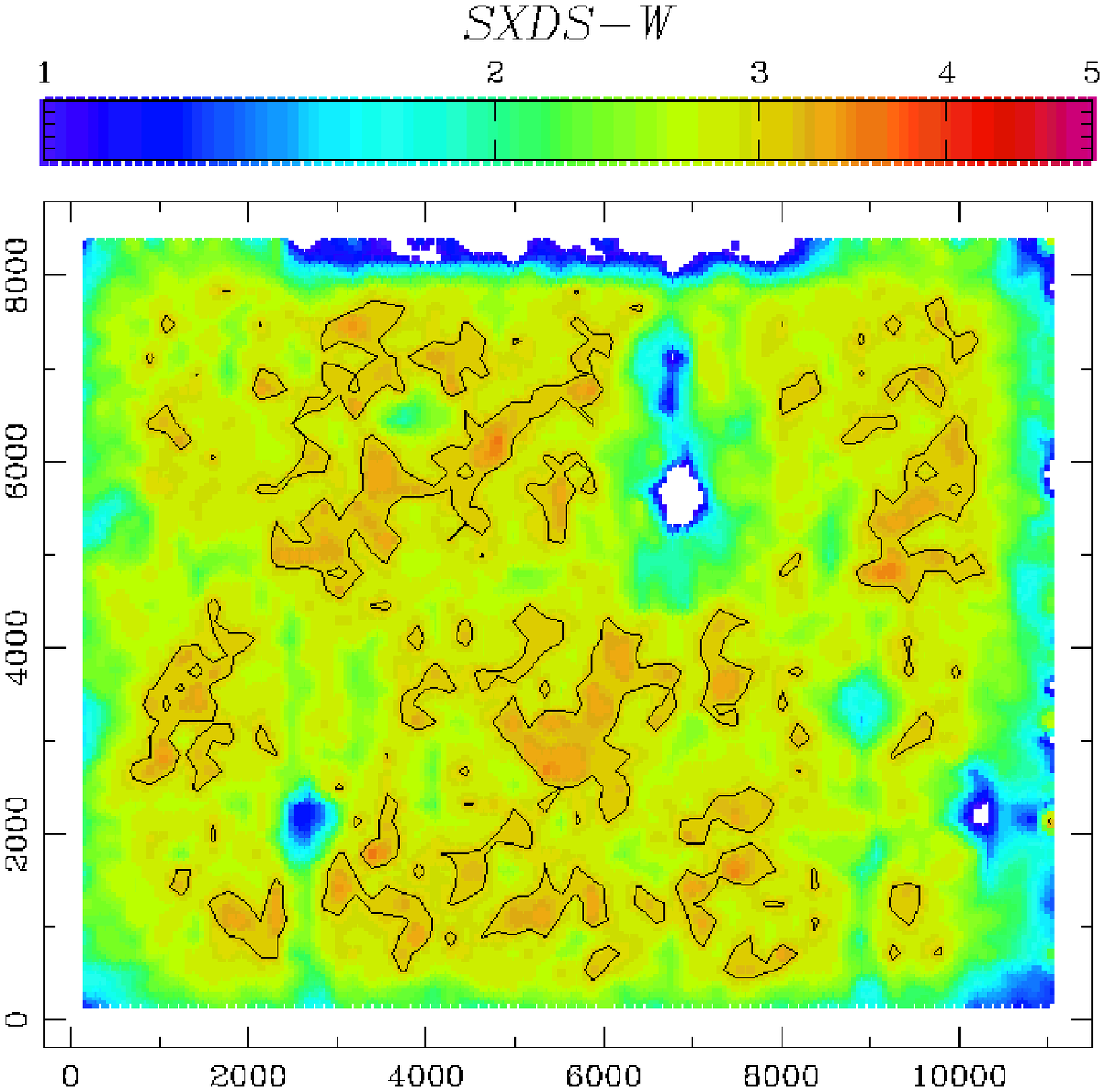}
\caption{Contour maps of the sky S/N distribution in the $R_c$ band. The
 deviation (in ADU) of sky background fluctuations per $2''$-diameter
 aperture are measured and converted to the S/Ns for objects with a
 magnitude of $27.5$ at each position.}
\label{fig:snmap}
\end{figure}

Figure~\ref{fig:snmap} shows the S/N distribution maps for 27.5-mag
objects thus obtained for each pointing.
The contour lines superimposed on the figures represent S/N=3 positions.
First, we see that the  brightest sources on the images significantly affect
the S/Ns around the sources.  These areas are flagged in the catalogs.
Second, each image shows almost an axisymmetric pattern of the S/N
distribution when we ignore the low S/N areas due to the brightest sources.
This pattern reflects the vignetting of the prime focus.  
Third, the pattern is slightly distorted for each image, probably due
to bright sources, vignetting by the auto-guider probe, difference in
the quantum efficiency among CCDs, and so on.  However, the difference
in S/Ns between in such distorted areas and in clean areas with higher
S/Ns is on the order of 20\% at the maximum. This difference is
negligible in most studies even for faint sources.  Thus, we can
securely utilize the data with quite homogeneous S/Ns across the field
of view.

\subsection{Completeness and False Detection Rate}\label{sec:comp_false}
Completeness of the object detection as a function of magnitude 
for each band and for each pointing is estimated.  The detection completeness is
determined based on a Monte Carlo simulation by adding artificial
objects which have Gaussian profiles with their Poisson photon noise
into a stacked image used to create the catalogs and then detect them
again in the same manner as in creating catalogs. The detection rate
thus obtained is regarded as the detection completeness.
Here the artificial objects are added to random positions in the effective
area of the image. The FWHMs of the Gaussian profiles are set to be the
same as the PSF sizes representative of the images which are derived
from Table~\ref{tab:stacked}. The artificial objects contaminated by any 
neighbor sources listed in the catalogs are removed in calculation of
the detection rate. If such blended objects are not removed from the
sample, the detection rate should slightly decrease. However, since the
contamination by neighbor objects leads to misidentification of the
artificial objects in the object detection process, we adopt the
approach which excludes the blended artificial objects, which was also
employed in Kashikawa et al. (2004).

\begin{figure}
\epsscale{1}
\plotone{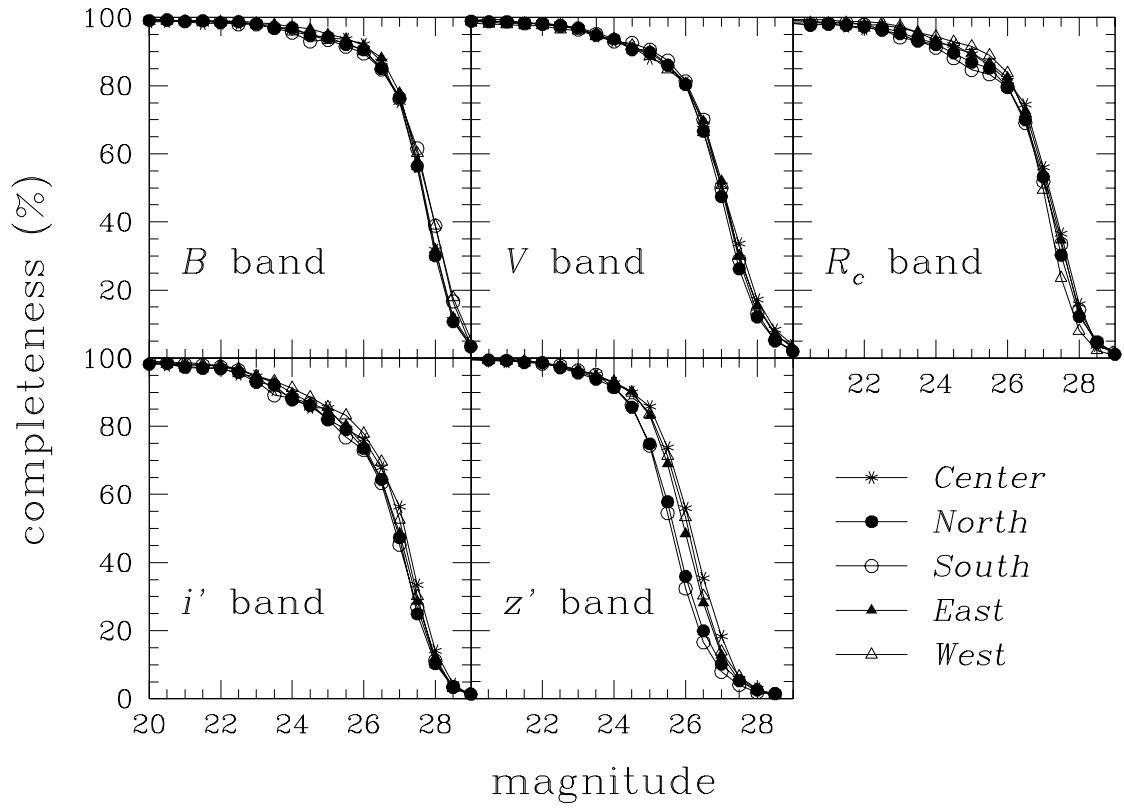}
\caption{Object detection completeness of the five pointings as a
 function of magnitude in each band.}
\label{fig:comp}
\end{figure}

The completeness determined by this procedure is shown in Figure~\ref{fig:comp}
for all of the five bands.  For the five pointings, the curves of the detection 
completeness drop similarly with increasing magnitude.  
We can summarize that the detection completeness in each image 
is at approximately $50\%$ for objects with the 5-$\sigma$ magnitude and
$30\%$ for objects with the 3-$\sigma$ magnitude, with a slight difference 
depending on the band. 

\begin{figure}
\epsscale{1}
\plotone{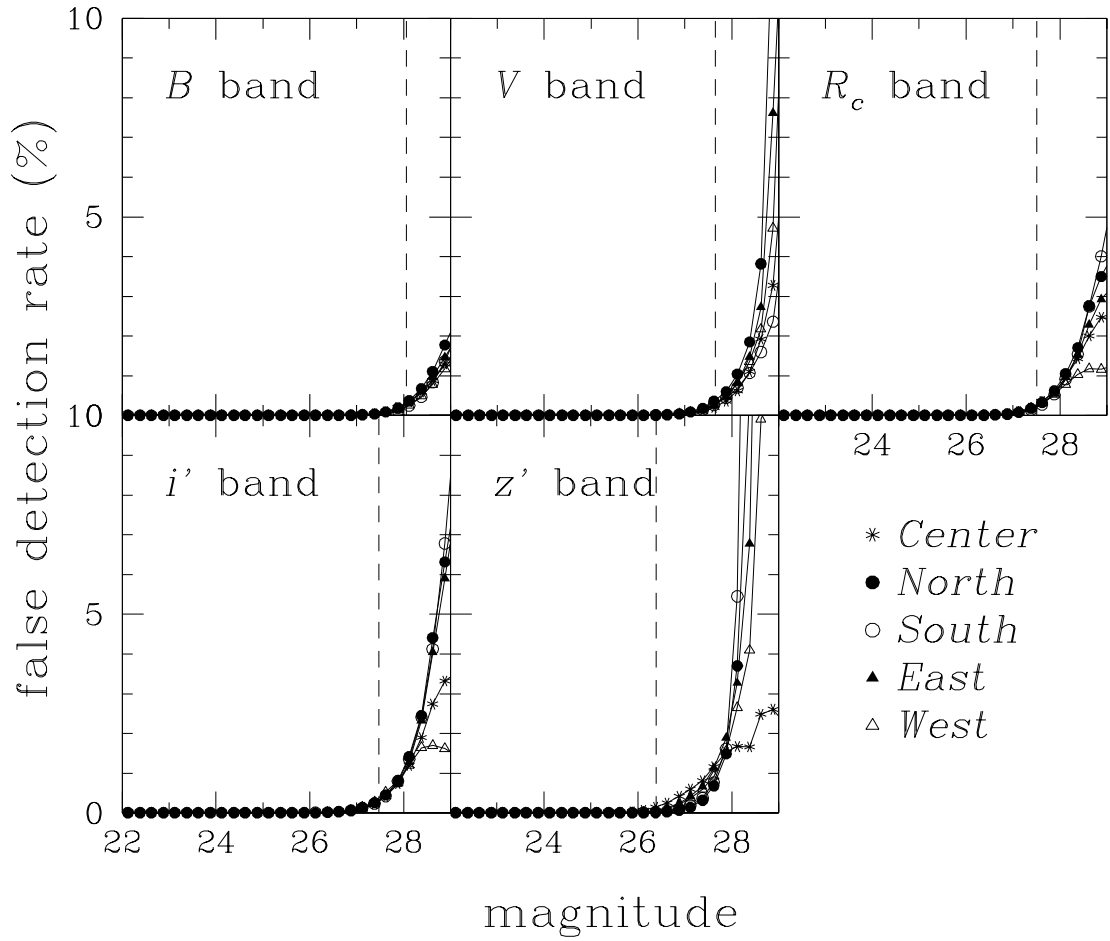}
\caption{False object detection rate for the five pointings as a
 function of magnitude in the five bands. The perpendicular dashed lines
 indicate the 3-sigma limiting magnitude measured in $2''$-diameter
 aperture in the center pointing ({\it SXDS-C}).}
\label{fig:false}
\end{figure}

On the other hand, rates of the false object detection are also
estimated as a function of magnitude. 
We generate negative images of each pointing and in each band by
multiplying all the counts of the stacked images by $-1$, and then perform
object detection for the negative images in the same manner as in creating
the catalogs.  
In this process, detected objects are considered to be spurious objects.
Thus we define the false detection rate as the number of the spurious objects 
divided by the total number of objects in the catalogs 
in each magnitude bin, i.e., $N_\mathrm{spurious}(m)/N_\mathrm{total}(m)$ (Figure~\ref{fig:false}).
It is seen that a contribution of the false detection is negligible, 
which does not exceed 0.5\% if any, in the magnitude range brighter
than the 3-$\sigma$ limiting magnitude.

\subsection{Magnitude Differences among the Catalogs}\label{sec:magdiff}
In the areas where two images overlap with each other, 
a large fraction of objects are detected and their magnitudes are
measured in both of the pointings. 
We can check systematic difference in magnitude of those objects
between the two catalogs.  
Since photometric zeropoints of the catalogs
are determined so that {\tt MAG\_AUTO} of objects in the overlapping
areas should coincide between any two catalogs (see
Section~\ref{sec:photo_calib}), the magnitude difference in {\tt MAG\_AUTO} is
negligible. {\tt MAG\_BEST} should be identical to {\tt MAG\_AUTO} for
isolated objects.
Therefore magnitude differences only for the $2''$- and $3''$-diameter aperture
magnitudes are investigated in the following manner.  

First, we choose only compact objects with FWHMs $<1.2$ arcsec 
which are listed in both the $z'$-selected catalog for {\it SXDS-C} and
that for another pointing which overlap with each other, then measure
the difference in magnitude of the objects by subtracting magnitude in the {\it SXDS-C}
catalog by that in the other.  Then, we determine the offset in the
magnitude difference, i.e., systematic magnitude difference, by fitting
the magnitude difference for a range of 21.0 to 23.5 mag by a least
square method.
Table~\ref{tab:diff_magap} lists the systematic magnitude difference
thus obtained for the $2''$- and $3''$- diameter aperture magnitudes in each
band for each pair of catalogs.
We note that $2''$- and $3''$-diameter aperture magnitudes indicate
small systematic differences of $\leq 0.05$ mag between the catalogs. 
 We think that this difference is probably due to difference in the PSF
 shape of the stacked images among the pointings. 

\begin{figure}
\epsscale{1}
\plotone{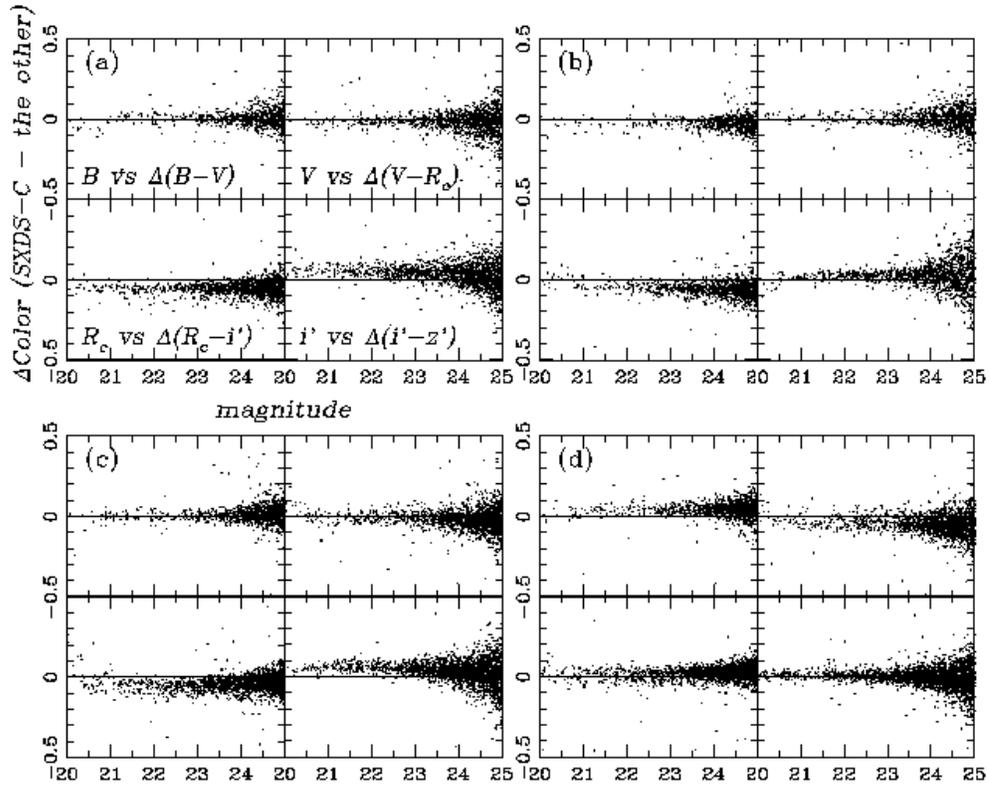}
\caption{Residuals in colors measured for the $2''$-diameter aperture
 magnitudes of the objects common in two separate pointing catalogs.}
\label{fig:colordiff}
\end{figure}

In order to see the effects of the systematic difference in the aperture
magnitudes, we plot differences in colors of objects between the
catalogs measured with the $2''$-diameter aperture magnitude
(Figure~\ref{fig:colordiff}). In this figure, for each of the four pairs 
of catalogs, the differences in colors ($B-V, V-R_c, R_c-i', i'-z'$) are
plotted.  We see the systematic difference in the $2''$-diameter aperture
colors of objects, which is no larger than $0.1$ mag, due to the difference
in the aperture magnitude between the catalogs.
Since the magnitudes in the catalogs are not adjusted for this
difference, the magnitude difference between the catalogs should be
taken into account in the case that the $2''$- or $3''$-diameter
aperture magnitude is used.  For {\tt MAG\_AUTO} and {\tt MAG\_BEST}, which are
considered as asymptotic total magnitude, the magnitude difference is 
negligible.

\begin{figure}
\epsscale{1}
\plotone{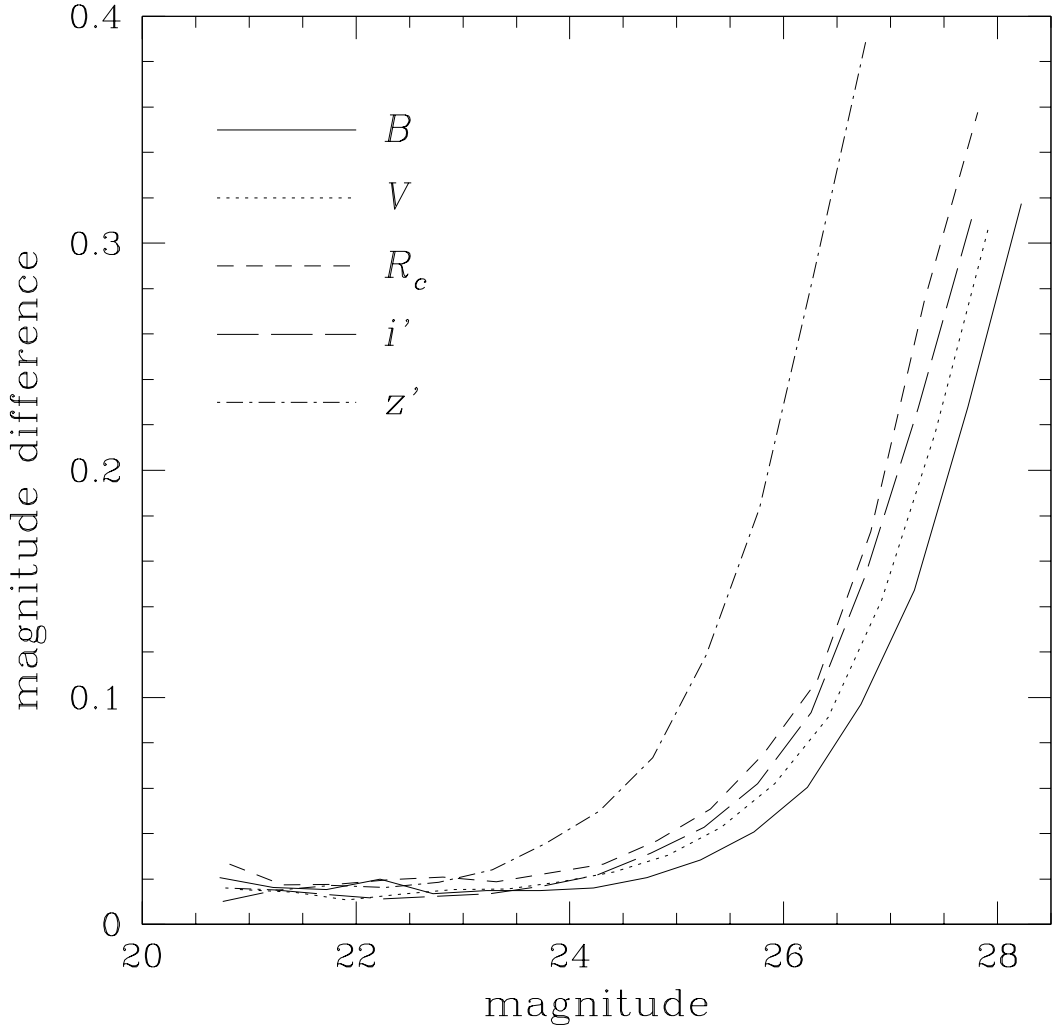}
\caption{Random magnitude difference as a function of magnitude in each
 band. The random magnitude differences are computed by comparing fluxes
 of compact objects with FWHMs $<1.2$ arcsec between the central
 pointing and the four surrounding pointings.}
\label{fig:magdiff_random}
\end{figure}

In the same manner as mentioned above, the r.m.s. of the magnitude
differences for each band are calculated as a function of {\tt
MAG\_AUTO}, which shows no systematic difference.  
Here the detection band for catalogs are chosen to be the same as the
band for which the magnitude difference is investigated. For instance, 
to obtain the magnitude difference in the $B$ band, the $B$-selected
catalogs are used.
Again, only the objects with FWHMs $<1.2$ arcsec are used for
measurement of the magnitude difference.
The r.m.s. of magnitude differences measured in the above way are shown
in Figure~\ref{fig:magdiff_random} for each band. It is seen that the
r.m.s. steadily increases with increasing magnitude.
The r.m.s. of the magnitude difference can be converted into random
magnitude error by dividing it by $\sqrt2$.
We find that the magnitude errors thus estimated are entirely consistent
with those expected from our calculation of the limiting magnitude
discussed in Section~\ref{sec:limmag}.
Note that the magnitude errors listed in our catalogs are computed 
by SExtractor assuming the simple Poisson statistics for the sky 
background fluctuations. 
The sky background noises measured in photometry apertures for the
stacked images are likely to be larger than those estimated by
SExtractor. Therefore the magnitude errors in the catalogs are
underestimates and are not suitable for immediate analysis such as SED 
fitting and photo-$z$ estimation.

\subsection{Number Counts of Galaxies}\label{sec:nm}
To examine the characteristics of the catalogs we compare the number
counts of galaxies with data from previous surveys.

We calculate the number counts of galaxies by correcting our
raw counts of detected objects against the detection completeness
estimated in Section~\ref{sec:comp_false}.
The detailed studies of density fluctuations of galaxies based on the
number counts or luminosity functions of galaxies can be 
found in elsewhere (e.g., Yamada et al. 2005). 
Star/galaxy separation is conducted based on FWHM and $2''$-diameter
fixed aperture magnitude of the objects, for which the effect of the
Galactic extinction (Table~\ref{tab:gal_ext}) is corrected.  The
criteria for the separation is determined so that both objects which
have sizes as same as PSF of the images and saturated objects are
effectively removed.
For fainter magnitudes ($>22.5-23$ mag), no separation is performed
since a large fraction of galaxies have small sizes similar to those of
stars, and contribution to the number count by stars is small at
the faint magnitudes.
The criteria for the separation adopted is summarized in Table~\ref{tab:sgsep}.

\begin{figure}
\epsscale{1}
\plotone{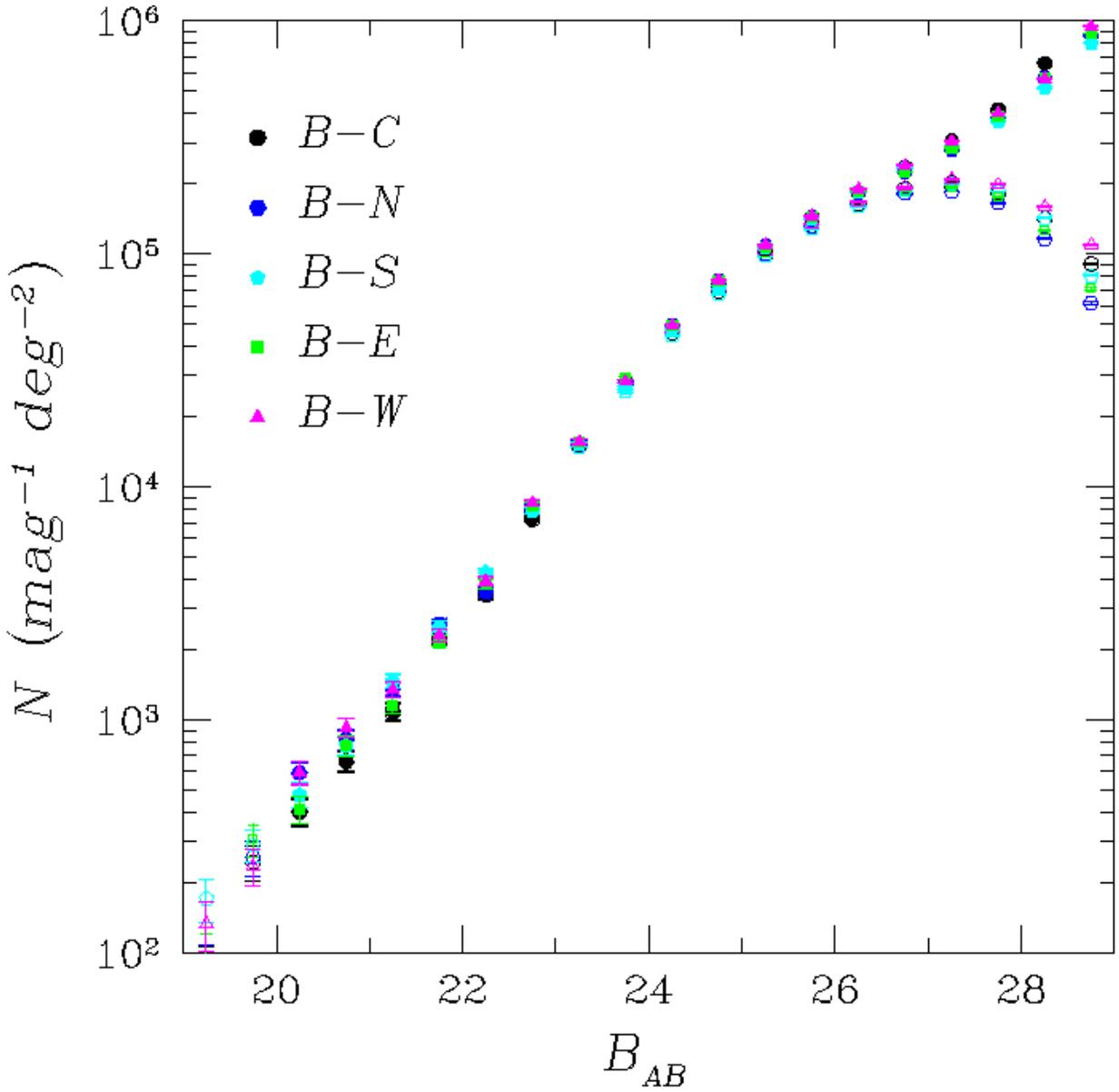}
\caption{$B$-band galaxy number counts for the five pointings. Open
 symbols are the raw numbers extracted using the star/galaxy separation
 process. Filled symbols are the numbers corrected for the detection
 completeness estimated with Gaussian artificial sources.}
\label{fig:nm_b}
\end{figure}

\begin{figure}
\epsscale{1}
\plotone{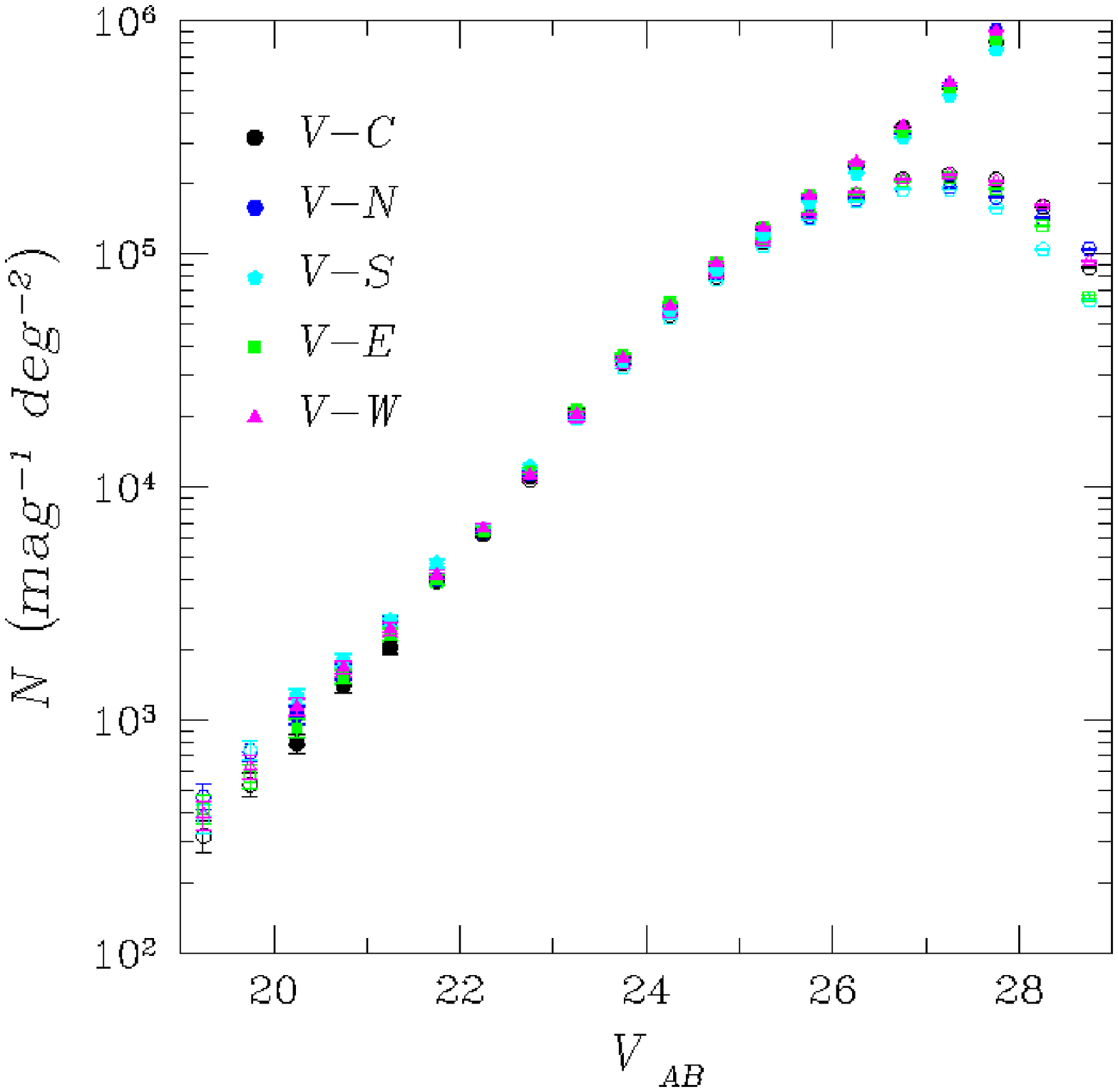}
\caption{$V$-band galaxy number counts for the five pointings. Symbols are the same as in Figure~\ref{fig:nm_b}.}
\label{fig:nm_v}
\end{figure}

\begin{figure}
\epsscale{1}
\plotone{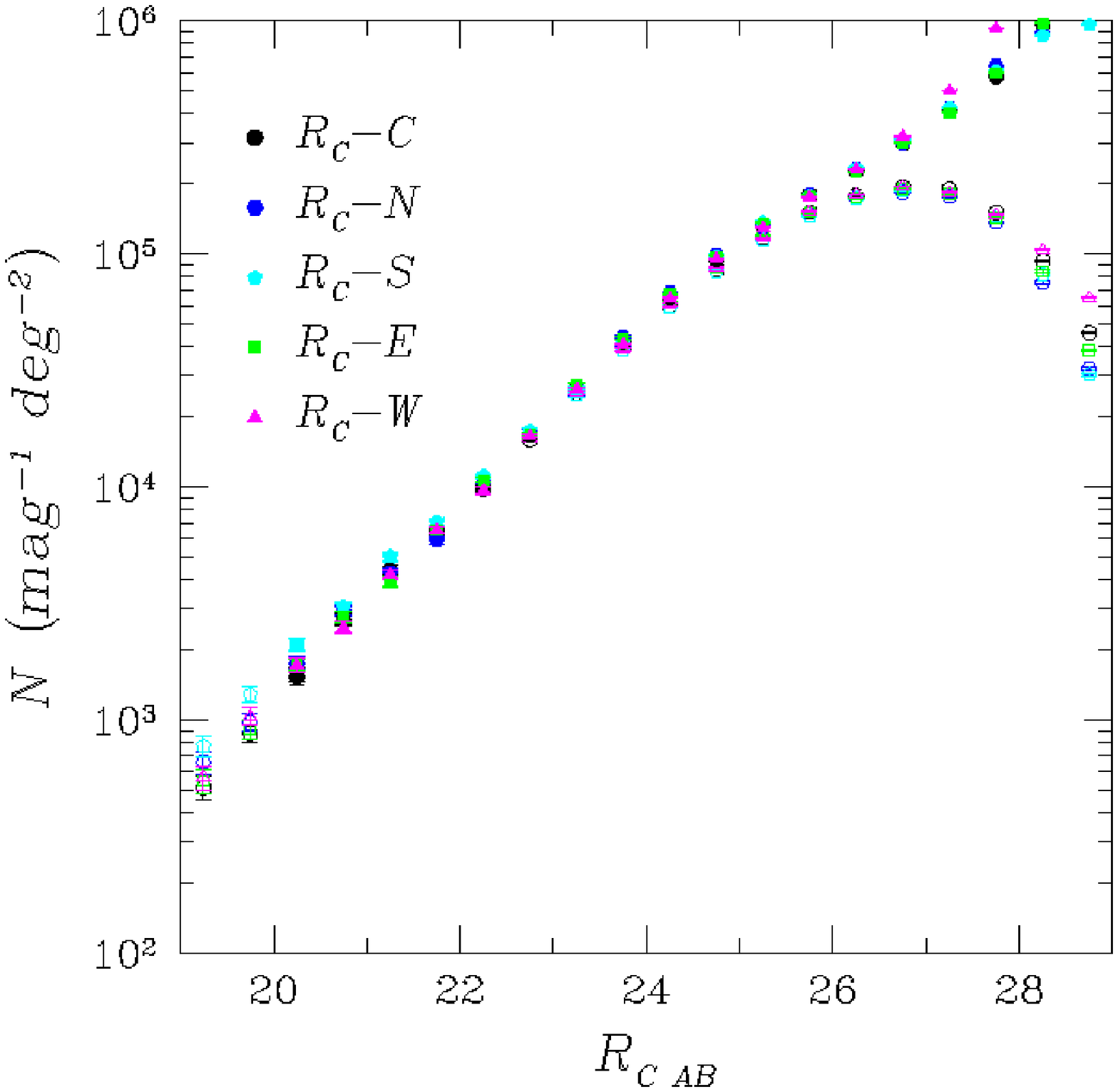}
\caption{$R_c$-band galaxy number counts for the five pointings. Symbols are the same as in Figure~\ref{fig:nm_b}.}
\label{fig:nm_r}
\end{figure}

\begin{figure}
\epsscale{1}
\plotone{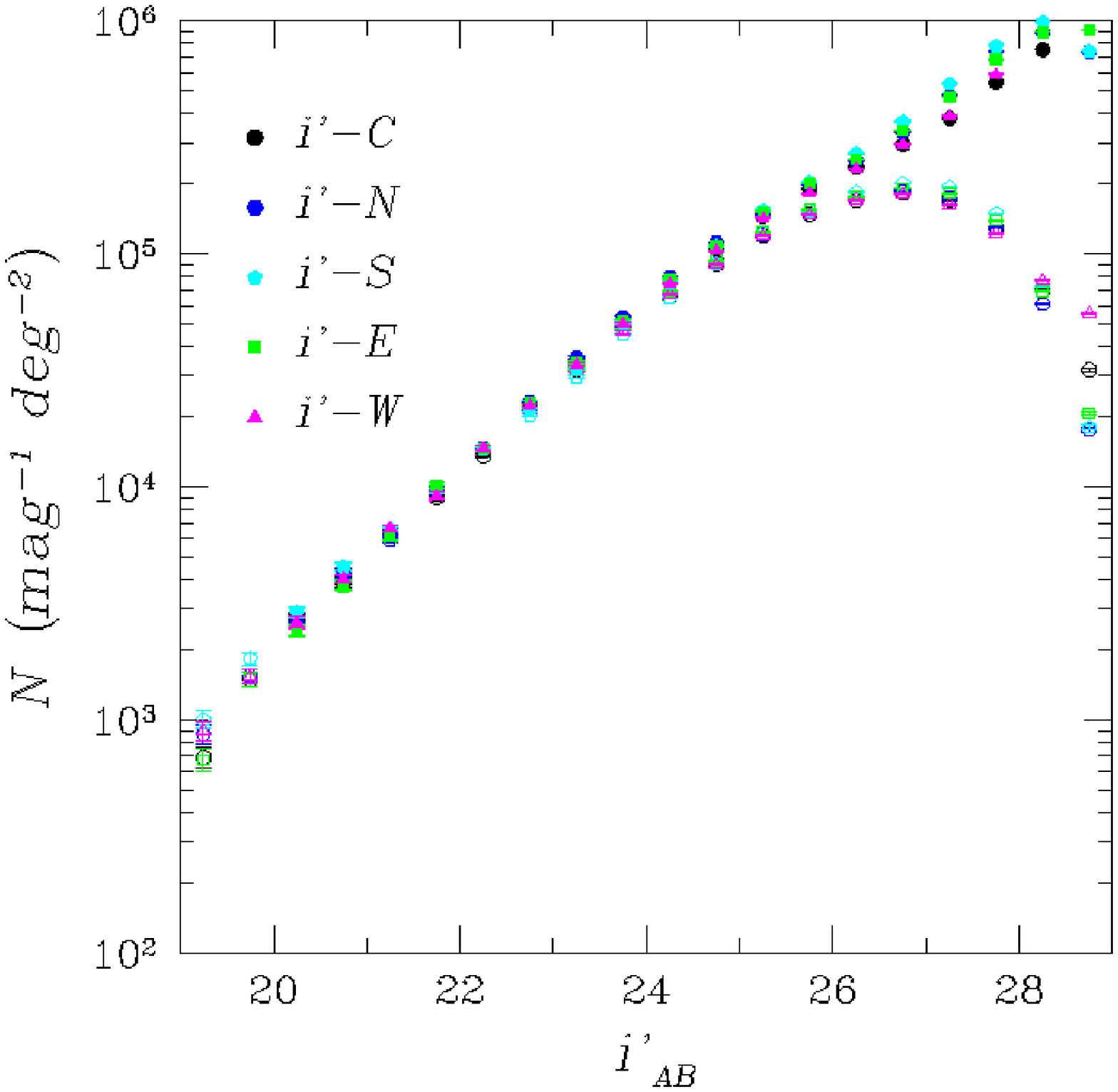}
\caption{$i'$-band galaxy number counts for the five pointings. Symbols are the same as in Figure~\ref{fig:nm_b}.}
\label{fig:nm_i}
\end{figure}

\begin{figure}
\epsscale{1}
\plotone{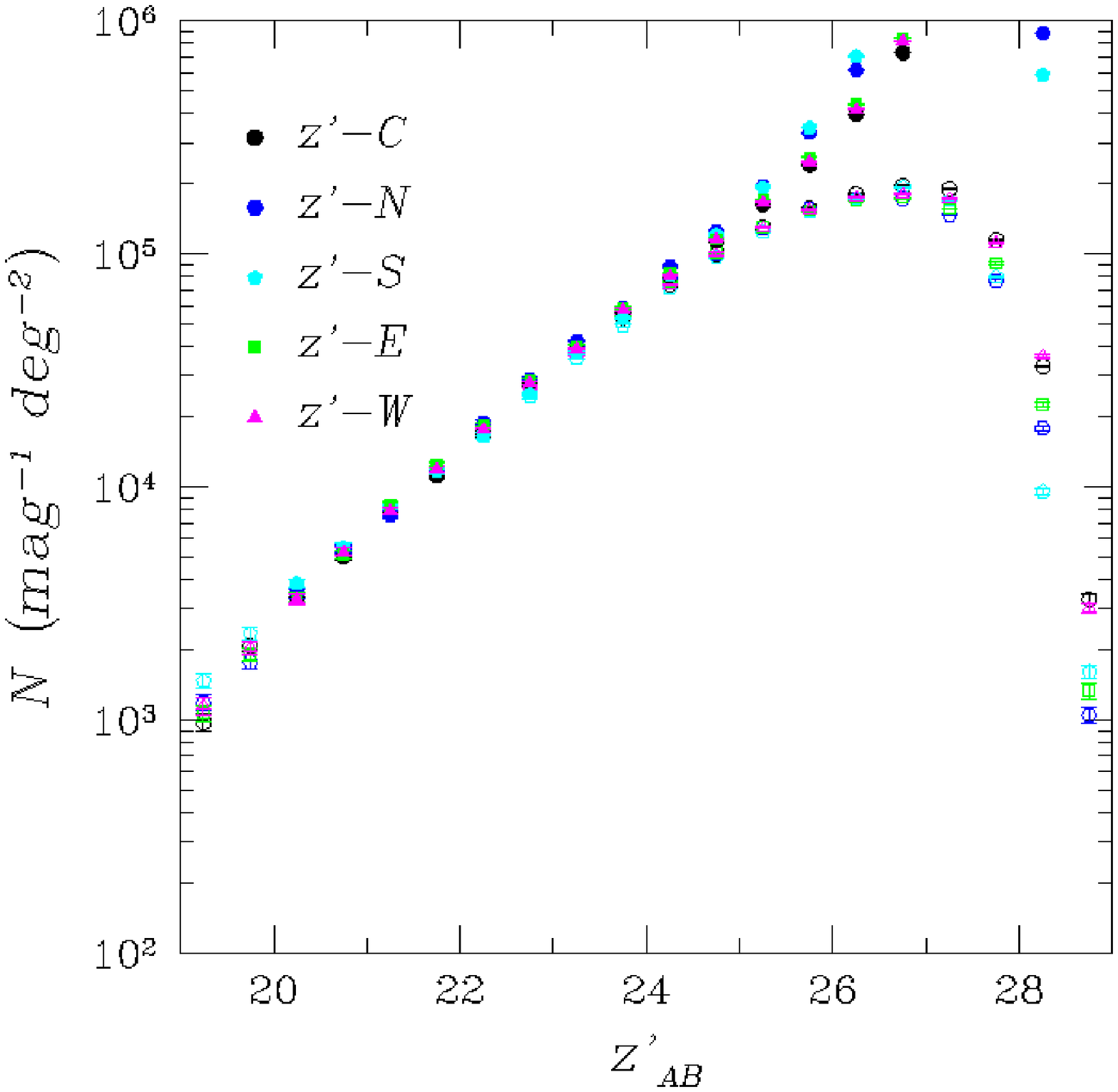}
\caption{$z'$-band galaxy number counts for the five pointings. Symbols are the same as in Figure~\ref{fig:nm_b}.}
\label{fig:nm_z}
\end{figure}

The results of the number counts of galaxies are as follows: 
First, internal comparison of the number counts among the five pointings
of the SXDS are shown for each band in Figure~\ref{fig:nm_b} to Figure~\ref{fig:nm_z}.
In each figure, raw number counts of the galaxy samples extracted by 
the star/galaxy separation process above are plotted with open symbols. 
Effective areas used to calculate the number counts per unit area are
given in Table~\ref{tab:stacked}.
The number counts of galaxies which are corrected for the detection
completeness are also superimposed with filled symbols.
Here, the correction is performed by using the detection completeness as
a function of magnitude which has been estimated for the Gaussian
profiles, i.e., 
$N(m)_\mathrm{corrected}=N(m)_\mathrm{observed}/ completeness(m)$.
Error bars are calculated on the assumption of simple Poisson noise. 
We can say that no large systematic difference is seen in 
the superposition of the number counts of galaxies among the five
pointings in $B, V$, and $R_c$ bands down to the 3-$\sigma$ magnitudes.  
Similarly, in the other two bands, the number counts are consistent 
with one another among the five pointings within the Poisson
errors at magnitudes brighter than $\sim 25$ mag.
However, differences in the number counts 
at fainter magnitudes are found by a factor of 1.4 in the $i'$ band and
1.7 in the $z'$ band. 
This might be partly due to the field-to-field variation in the number
counts of galaxies. However, since we have used only the Gaussian profiles
with FWHM$\sim 0.8$ arcsec in the simulation of the detection completeness, 
the completeness is likely to be overestimated at a 
faint magnitude range where most objects have extended profiles.
So, we cannot conclude that the differences in the number counts at 
the faint end seen in the $i'$ and $z'$ bands indicates the
field-to-field variation.

\begin{figure}
\epsscale{1}
\plotone{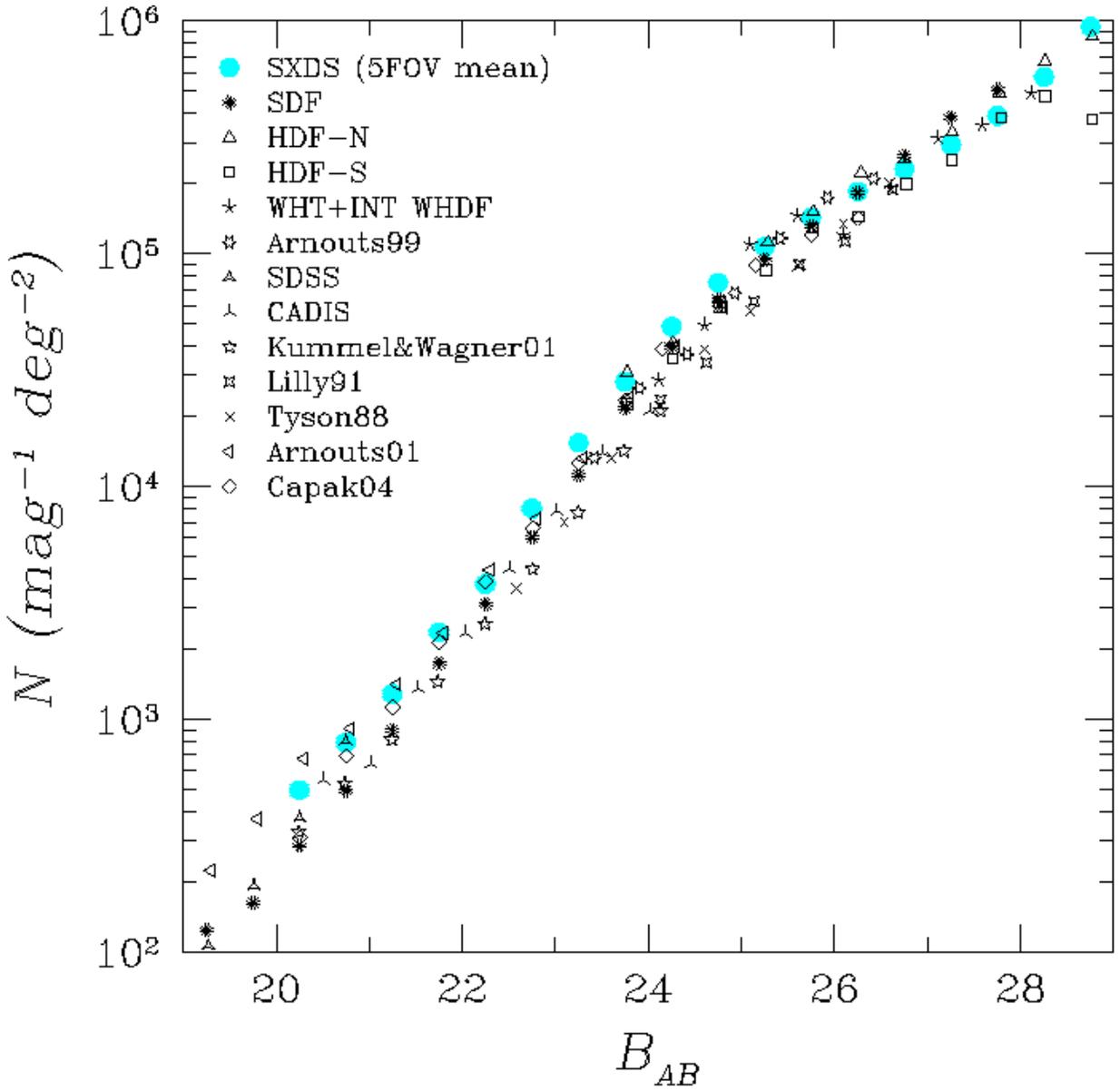}
\caption{Comparison of the galaxy number count of the SXDS (blue filled
 circles) with other major optical imaging surveys. The SXDS values are
 corrected for the detection completeness and the mean value for the
 five pointings is plotted.}
\label{fig:nm_mean_b}
\end{figure}

\begin{figure}
\epsscale{1}
\plotone{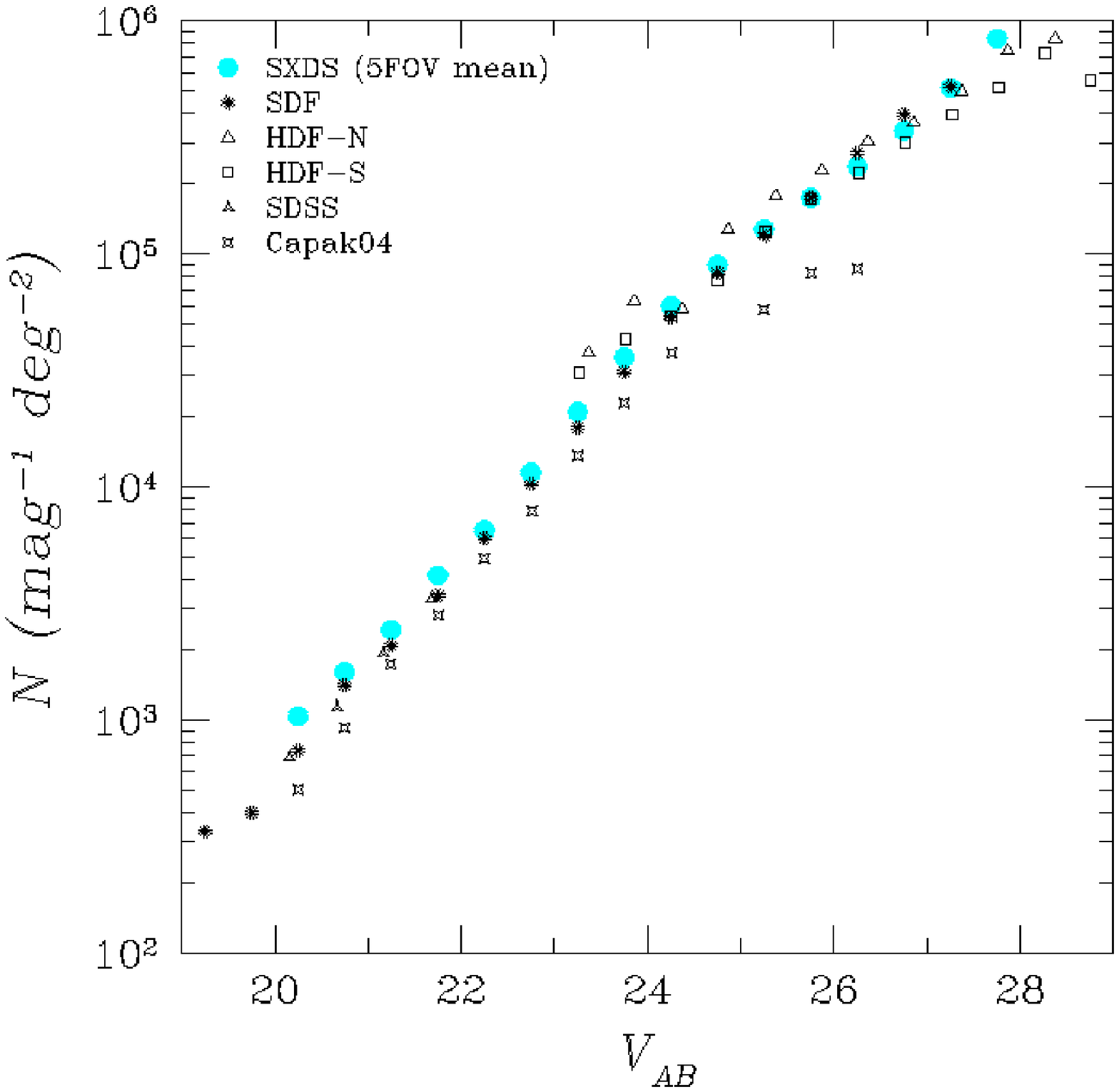}
\caption{Same as Figure~\ref{fig:nm_mean_b}, but for $V$-band data.}
\label{fig:nm_mean_v}
\end{figure}

\begin{figure}
\epsscale{1}
\plotone{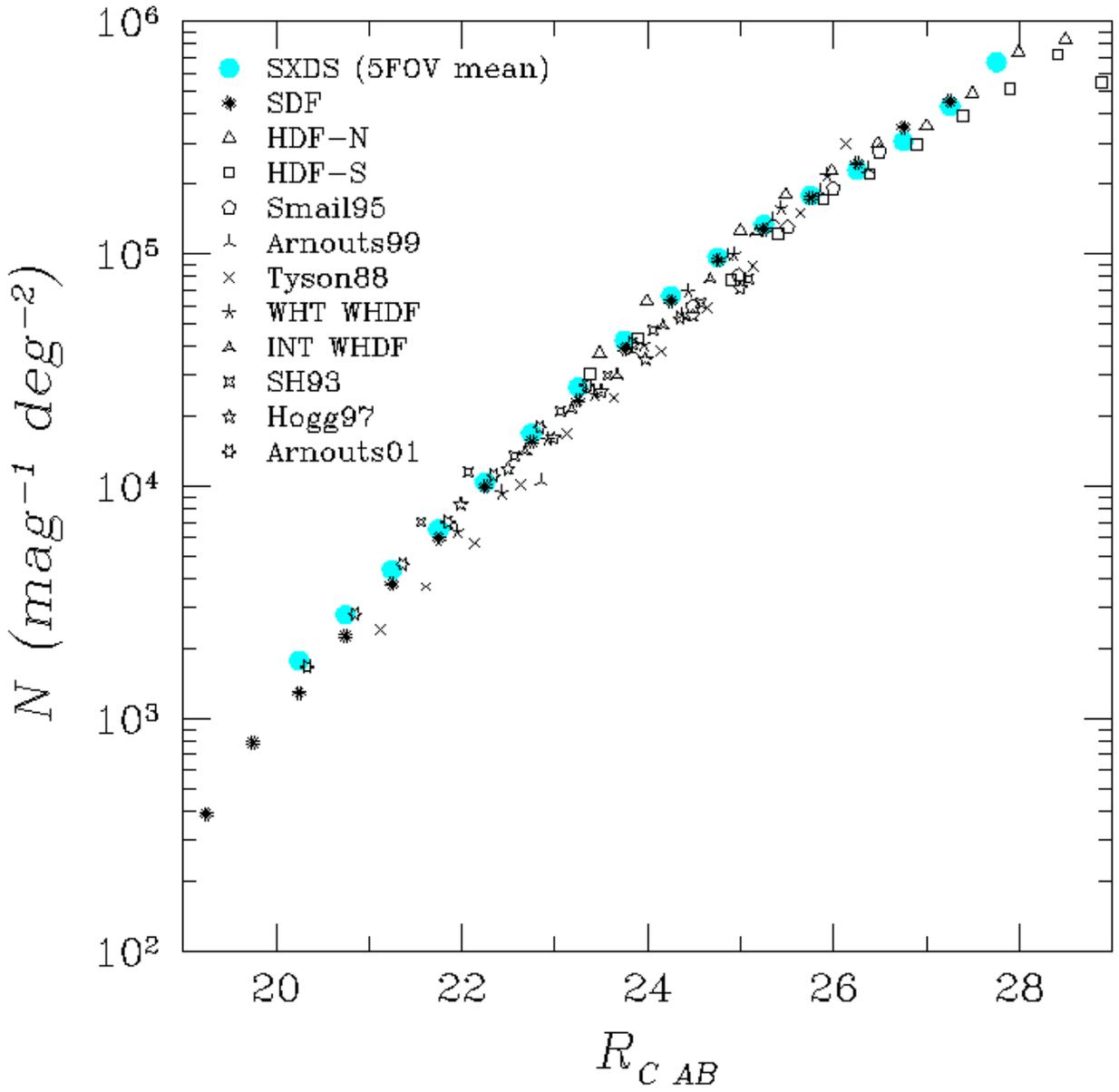}
\caption{Same as Figure~\ref{fig:nm_mean_b}, but for $R_c$-band data.}
\label{fig:nm_mean_r}
\end{figure}

\begin{figure}
\epsscale{1}
\plotone{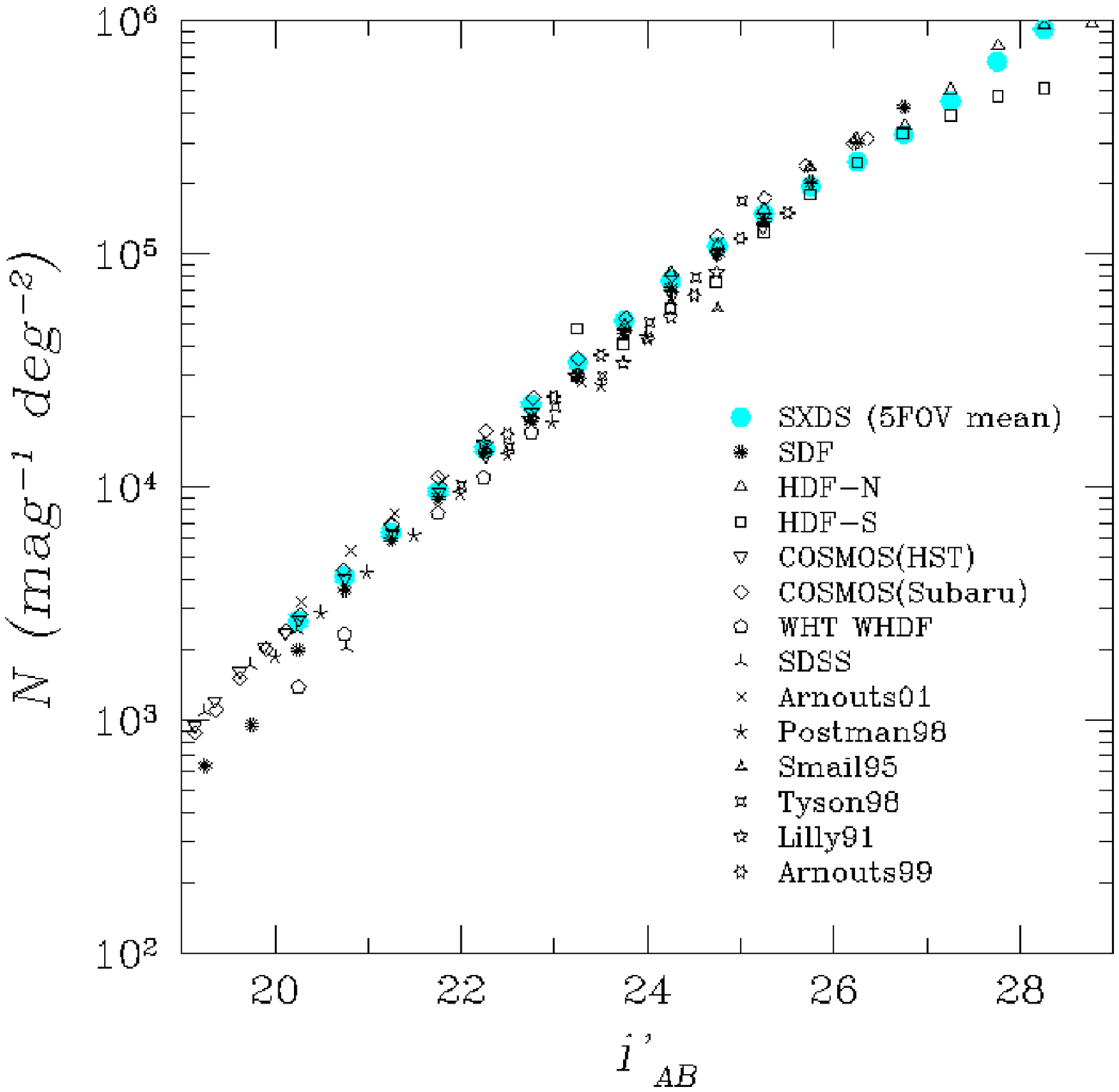}
\caption{Same as Figure~\ref{fig:nm_mean_b}, but for $i'$-band data.}
\label{fig:nm_mean_i}
\end{figure}

\begin{figure}
\epsscale{1}
\plotone{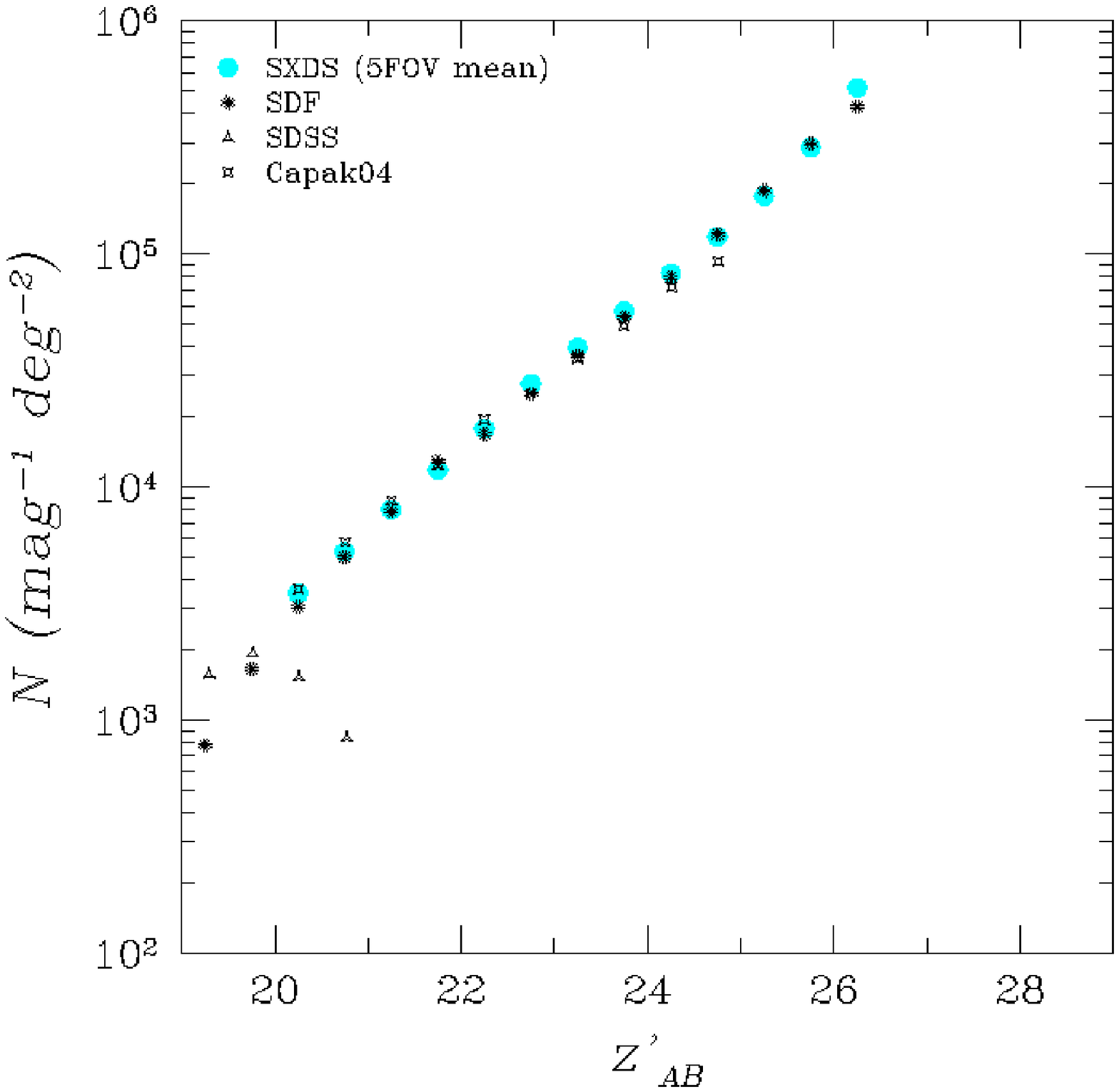}
\caption{Same as Figure~\ref{fig:nm_mean_b}, but for $z'$-band data.}
\label{fig:nm_mean_z}
\end{figure}

Next, the number counts in the SXDF are compared with results by previous
surveys (Figure~\ref{fig:nm_mean_b} to Figure~\ref{fig:nm_mean_z}).  
Here, we calculate mean number counts of galaxies corrected for the
detection completeness for the entire SXDF using the above sample for
the five pointings, and plot them with filled circles with error
bars. The error bars are again based on the Poisson noise. 
The results in other surveys are derived from Kashikawa et al. (2004),
Capak et al. (2007), and Metcalfe (2007)\footnote{\tt http://star-www.dur.ac.uk/\~{}nm/pubhtml/counts/counts.html} and the references cited therein.
A complete list of references compared in the figures is as follows:
SDF -- Kashikawa et al. 2004, HDFs/WHDFs -- Metcalfe et al. 2001,
Arnouts99 -- Arnouts et al. 1999, SDSS -- Yasuda et al. 2001, CADIS -- Huang
et al. 2001, Kummel\&Wagner01 -- K\"{u}mmel \& Wagner 2001, Tyson88 -- Tyson 
1988, Arnouts01 -- Arnouts et al. 2001, Capak04 -- Capak et al. 2004,
Smail95 -- Smail et al. 1995, SH93 -- Steidel \& Hamilton 1993, Hogg97 -- Hogg 
et al. 1997, COSMOS -- Capak et al. 2007, Postman98 -- Postman et al. 1998,
Lilly91 -- Lilly, Cowie, \& Gardner 1991.

From the comparison of the number counts among various surveys, 
we understand that our sample is consistent with results in other
surveys for blank fields for the following reasons.  First, the mean 
number counts of the SXDS show a good agreement with those from
previous surveys down to magnitude where the number counts start to turn off.  
Second, it reproduces the faint-end data points
derived from Hubble Deep Field (HDF). Again, the detection completeness 
simulated with the Gaussian profiles might cause a slightly large
uncertainty of the corrected number counts at the faint end in each band.
Thus, we conclude that the SXDS catalogs in the present study have
characteristics similar to those in previous surveys and can be applied
to general statistical studies on faint objects ranging from Galactic
objects to high-redshift galaxies. 

\begin{figure}
\epsscale{1}
\plotone{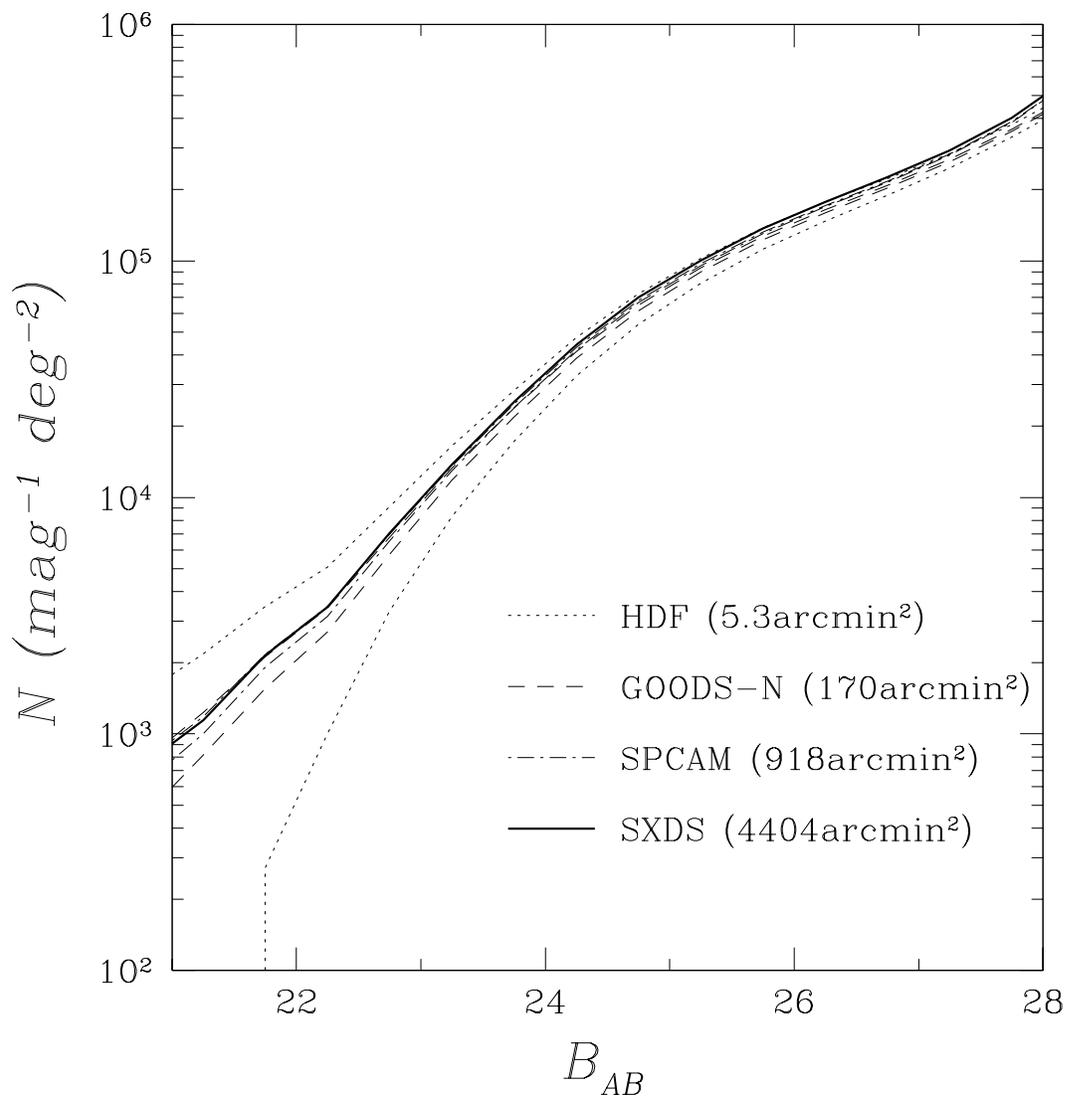}
\caption{$B$-band number count uncertainties of galaxies, due to the
 Poisson error, for the Hubble Deep Field (HDF), GOODS North field
 (GOODS-N), a single Suprime-Cam field (SPCAM) are plotted with the SXDS
 number count (solid line). The uncertainty for each survey is
 represented by a pair of lines plotted in the figure.}
\label{fig:poisson}
\end{figure}

We would like to mention how the Poisson error on the number counts is
decreased by increase of the survey area.
Figure~\ref{fig:poisson} shows a comparison of the Poisson error 
associated with the number counts of galaxies among three surveys which
are based on different survey areas, namely the HDF-North (5.3 arcmin$^2$),
GOODS-North (170 arcmin$^2$) and Suprime-Cam 1 FOV (918 arcmin$^2$).  The
uncertainty in the number counts due to the Poisson error is shown
with a pair of lines for each of the surveys.  For reference, the number
count of SXDS is superimposed with a thick solid line.
From the figure, we clearly see that the Poisson error decreases with
increasing survey area.  In particular, at the bright magnitudes of
$21-23$ mag, a large uncertainty by a factor of $>3$ in the number
count is found for the HDF-North area (dotted lines). For the same
magnitude range, the uncertainty for a field of view of Suprime-Cam
(dot-and-dash lines) is $8-15$ times as small as that for the HDF-North
area and $1.5-2.5$ times as small as that for the GOODS-North area.
Even at the faintest magnitudes of $>27$ mag, a significant difference
in the uncertainty from one to another survey area is seen.
Thus it is emphasized that the SXDS data, which consists of five 
pointings of Suprime-Cam field of view, is a useful data set for studies
on celestial objects which dominate a large fraction of the number
counts without suffering from the Poisson error and the field-to-field
variation as well, e.g., a study of evolution of the LSSs. 

\section{SUMMARY}  
The optical imaging observations of the SXDS project are carried out using 
the Suprime-Cam on Subaru Telescope in $B, V, R_c, i'$ and $z'$ bands.
The SXDF has a contiguous area coverage of $\sim$1.2 deg$^{2}$, which
consists of five Suprime-Cam pointings.  

The photometric zeropoints are determined with an absolute accuracy of
no larger than 0.05 mag r.m.s. in the photometry.
The r.m.s. of the astrometric accuracies across the field is
on the order of 0.2 arcsec in both RA and Dec.  The systematic
differences in the adopted world coordinates between different pointings
are within about 1 arcsec at the outer edge of the field of view. Thus,
our SXDS catalogs have a good enough positional accuracies to perform
follow-up spectroscopy. 

Multi-waveband photometric catalogs of detected objects are created for
each band and for each pointing. Each of the catalogs contains more than
a hundred sixty thousand of objects.  The catalogs have quite
homogeneous S/Ns across the field.
The achieved limiting magnitudes in each band are $B=28.4, V=27.8,
R_{c}=27.7, i'=27.7$ and $z'=26.6$ ($AB,  3\,\sigma, \phi = 2''$).  

The detection completeness as a function of magnitude is estimated by 
a Monte-Carlo simulation assuming a Gaussian profile. The number counts
of galaxies in the SXDF in each band are computed by correcting
the detection completeness, which are consistent with each other among
the five pointings with a slight difference at faint magnitudes. The
mean number counts of galaxies averaged over the five pointings show
a good agreement with results from previous surveys down to the faint-end 
magnitude. 

With the aid of the wide coverage area of SXDS data, the uncertainty of
the number counts of galaxies due to the Poisson error is greatly decreased.
It is emphasized that the SXDS data is extremely useful for pursuing 
studies on celestial objects spreading in a wide field without suffering
from the Poisson error and the field-to-field variation. 
The SXDS catalogs can be applied to studies, scientific objectives of
which range from the Galactic objects to the large scale structures of
the universe.

The optical data, the compiled photometric catalogs, and configuration 
files used to create the catalogs have been released to the
public and can be retrieved from the public data archives server of National 
Astronomical Observatory of Japan, {\tt http://www.naoj.org/Science/SubaruProject/SXDS/index.html}.

\smallskip

Acknowledgments

We thank Dr. Masafumi Yagi for his kind suggestions in analyzing the data based
on the software package developed by himself. We thank Dr. Nobunari Kashikawa for
fruitful discussions. The anonymous referee would also be appreciated for his/her 
constructive suggestions.
We would like to convey our gratitude to the Subaru Telescope
builders and staff for their invaluable help to complete the observing
run with Suprime-Cam. 
Data reduction/analysis in this work was in part carried out on "sb"
computer system operated by the Astronomical Data Center (ADC) and
Subaru Telescope of the National Astronomical Observatory of Japan. Use
of the UH 2.2-m telescope for the observations is supported by NAOJ.

%
%

\clearpage

\clearpage

\clearpage

\clearpage

\clearpage

\clearpage

\clearpage

\clearpage

\clearpage

\clearpage

\clearpage

\clearpage


\clearpage

\clearpage

\clearpage
\begin{deluxetable}{cccc}
\tablecaption{COORDINATES OF SXDS FIVE POINTINGS\label{tab:coords}}
\tablehead{
 \colhead{Pointing} & 
 \colhead{RA (J2000)} &  
 \colhead{Dec (J2000)} &
 \colhead{Position Angle of FOV} 
}
\startdata
Center & $02^h18^m00^s.000$ & $-05\degr00'00.''00$ & North is up \\
North & $02^h18^m00^s.000$ & $-04\degr35'00.''00$  & North is up \\
South & $02^h18^m00^s.000$ & $-05\degr25'00.''00$ & North is up \\
East  & $02^h19^m47^s.070$ & $-05\degr00'00.''00$ & East is up \\
West  & $02^h16^m12^s.930$ & $-05\degr00'00.''00$ & West is up \\
\enddata
\tablecomments{Field names for each pointing and their positional
 coordinates.  PA, the orientation of the FOV on the sky are chosen so
 that the overhead time to rotate the position angle are minimized and 
 effective area with high signal-to-ratios are is maximized.}
\end{deluxetable}

\clearpage
\begin{deluxetable}{cccccclccccccl}
\rotate
\tabletypesize{\scriptsize}
\tablecaption{SUMMARY OF OBSERVATIONS AND DATA OF THE SXDS\label{tab:log}}
\tablehead{
 \colhead{} &  
 \colhead{Exp. Time} &  
 \colhead{PSF Size} & 
 \colhead{Zeropts for Images} & 
 \colhead{$m_{lim}$} & 
 \colhead{} &
 \colhead{}\\
 \colhead{Band-Pointing} &  
 \colhead{(min)} &  
 \colhead{(arcsec)} & 
 \colhead{(mag$_{\rm AB}$ADU$^{-1}$)} & 
 \colhead{(mag$_{\rm AB}$)} & 
 \colhead{$N_{\rm obj}$} &
 \colhead{Date of Observations} 
}
\startdata
{\it $B$-C}    & 345 & 0.80 & 34.723 & 28.09 & 197,317 & {2002, Sep. 29/30, Oct. 01} \\
{}&{}&{}&{}&{}&{}&{2003, Nov. 17} \\
{\it $B$-N}    & 330 & 0.84 & 34.701 & 28.39 & 176,372 & {2002, Nov. 02/05/27, Dec. 01/07}\\
{}&{}&{}&{}&{}&{}&{2003, Nov. 17} \\
{\it $B$-S}    & 330 & 0.82 & 34.706 & 28.33 & 189,916 & {2002, Nov. 02/04/05/27, Dec. 01/07}\\
{}&{}&{}&{}&{}&{}&{2003, Nov. 17} \\
{\it $B$-E}    & 330 & 0.82 & 34.698 & 28.06 & 179,478 & {2002, Nov. 04/05, Dec. 01/07}\\
{}&{}&{}&{}&{}&{}&{2003, Nov. 17} \\
{\it $B$-W}    & 330 & 0.78 & 34.716 & 28.21 & 197,770 & {2002, Nov. 04/05/27, Dec. 01/07}\\
{}&{}&{}&{}&{}&{}&{2003, Nov. 17} \\
{}&{}&{}&{}&{}&{}&{} \\
{\it $V$-C}    & 319 & 0.72 & 33.639 & 27.78 & 213,851 & {2003, Oct. 02/24/26} \\
{}&{}&{}&{}&{}&{}&{2004, Jan. 16, Oct. 09, Dec. 10/14/15} \\
{}&{}&{}&{}&{}&{}&{2005, Jan. 05} \\
{\it $V$-N}    & 313 & 0.80 & 33.648 & 27.65 & 203,299 & {2003, Oct. 02/26, Nov. 17}\\
{}&{}&{}&{}&{}&{}&{2004, Jan. 16, Dec. 10/14/15} \\
{}&{}&{}&{}&{}&{}&{2005, Jan. 05, Sep. 28} \\
{\it $V$-S}    & 321 & 0.82 & 33.643 & 27.75 & 186,545 & {2003, Oct. 02/24/26}\\
{}&{}&{}&{}&{}&{}&{2004, Jan. 16, Dec. 10/14/15} \\
{}&{}&{}&{}&{}&{}&{2005, Jan. 05, Sep. 28} \\
{\it $V$-E}    & 291 & 0.76 & 33.639 & 27.77 & 192,951 & {2003, Oct. 02/26}\\
{}&{}&{}&{}&{}&{}&{2004, Jan. 16, Dec. 10/14/15} \\
{}&{}&{}&{}&{}&{}&{2005, Jan. 05, Sep. 28} \\
{\it $V$-W}    & 293 & 0.72 & 33.649 & 27.72 & 205,915 & {2003, Oct. 02/26, Nov. 17}\\
{}&{}&{}&{}&{}&{}&{2004, Jan. 16, Dec. 10/14/15} \\
{}&{}&{}&{}&{}&{}&{2005, Jan. 05, Sep. 28} \\
{}&{}&{}&{}&{}&{}&{} \\
{\it $R_c$-C}  & 248 & 0.76 & 34.315 & 27.57 & 188,079 & {2002, Sep. 30, Oct. 01/07}\\
{}&{}&{}&{}&{}&{}&{2003, Nov. 17} \\
{\it $R_c$-N}  & 232 & 0.78 & 34.276 & 27.74 & 174,365 & {2002, Nov. 01/09/30, Dec. 06}\\
{}&{}&{}&{}&{}&{}&{2003, Nov. 17} \\
{\it $R_c$-S}  & 232 & 0.74 & 34.219 & 27.67 & 178,416 & {2002, Nov. 01/09/30, Dec. 06}\\
{}&{}&{}&{}&{}&{}&{2003, Nov. 17} \\
{\it $R_c$-E}  & 232 & 0.76 & 34.259 & 27.51 & 173,586 & {2002, Nov. 02/09/30, Dec. 06}\\
{}&{}&{}&{}&{}&{}&{2003, Nov. 17} \\
{\it $R_c$-W}  & 232 & 0.82 & 34.247 & 27.53 & 186,648 & {2002, Nov. 02/09/30, Dec. 06}\\
{}&{}&{}&{}&{}&{}&{2003, Nov. 17} \\
{}&{}&{}&{}&{}&{}&{} \\
{\it $i'$-C}   & 647 & 0.78 & 34.055 & 27.62 & 181,352 & {2002, Sep. 29/30, Nov. 01/02/05/09/27/29, Dec. 06/07}\\
{}&{}&{}&{}&{}&{}&{2003, Oct. 20/21} \\
{\it $i'$-N}   & 440 & 0.76 & 34.042 & 27.66 & 176,394 & {2002, Sep. 29/30, Nov. 01/02/09/29, Dec. 07}\\
{}&{}&{}&{}&{}&{}&{2003, Sep. 22, Oct. 02/21} \\
{\it $i'$-S}   & 309 & 0.68 & 34.046 & 27.47 & 187,791 & {2002, Sep. 29/30, Nov. 01/02/09/29}\\
{}&{}&{}&{}&{}&{}&{2003, Sep. 22, Oct. 02} \\
{\it $i'$-E}   & 368 & 0.74 & 33.986 & 27.49 & 175,404 & {2002, Sep. 29/30, Nov. 01/02/09/29, Dec. 07}\\
{}&{}&{}&{}&{}&{}&{2003, Sep. 22, Oct. 02/21} \\
{\it $i'$-W}   & 598 & 0.82 & 34.087 & 27.58 & 178,543 & {2002, Sep. 29/30, Nov. 01/02/05/09/27/29, Dec. 06/07}\\
{}&{}&{}&{}&{}&{}&{2003, Oct. 20/21} \\
{}&{}&{}&{}&{}&{}&{} \\
{\it $z'$-C}   & 217 & 0.70 & 33.076 & 26.57 & 183,324 & {2002, Sep. 29/30, Oct. 01, Nov. 04/05} \\
{\it $z'$-N}   & 252 & 0.74 & 32.278 & 26.64 & 163,324 & {2002, Nov. 04/05/10}\\
{}&{}&{}&{}&{}&{}&{2003, Sep. 21/22, Oct. 01/02} \\
{\it $z'$-S}   & 184 & 0.76 & 32.258 & 26.39 & 167,779 & {2002, Nov. 04/05/10}\\
{}&{}&{}&{}&{}&{}&{2003, Sep. 21/22, Oct. 01} \\
{\it $z'$-E}   & 267 & 0.74 & 32.743 & 26.49 & 160,415 & {2002, Nov. 04/05/10}\\
{}&{}&{}&{}&{}&{}&{2003, Sep. 21/22, Oct. 01/02} \\
{\it $z'$-W}   & 311 & 0.74 & 32.776 & 26.60 & 167,748 & {2002, Nov. 04/05/10}\\
{}&{}&{}&{}&{}&{}&{2003, Sep. 21/22, Oct. 01/02} \\
\enddata
\tablecomments{Explanations of columns: (column 1) Filter pass-band ($B, V, Rc, i'$, or $z'$) and the pointed field (center = C, north = N, south =S, east = E, and west =W). (2) Total exposure time in minutes. (3) The PSF size of the stacked image. (4) Photometric zeropoint. (5) A 3-sigma limiting magnitude in AB magnitude, measured by random sampling with $2''$-diameter aperture. (6) Number of objects detected in the image. (7) Date of observations.}
\end{deluxetable}

\clearpage
\begin{deluxetable}{cccc}
\tablecaption{IMAGE SIZE AND THE AREA COVERED BY THE SXDS STACKED IMAGES\label{tab:stacked}}
\tablehead{
 \colhead{}          & \colhead{PSF Size} & \colhead{Area Covered} & \colhead{Effective Area} \\
 \colhead{Pointing}  & \colhead{(arcsec)} & \colhead{(arcmin${}^2$)} & \colhead{(arcmin${}^2$)} 
}
\startdata
Center (C) & 0.80 & 1041.9 & 968.4  \\
North (N) & 0.84 & 1011.7 & 951.8  \\
South (S) & 0.82 & 1008.4 & 967.3  \\
East  (E) & 0.82 & 986.4  & 917.3  \\
West  (W) & 0.82 & 1017.7 & 913.9  \\
Total  & ... & 4404.4 & 4057.0  \\
\enddata
\tablecomments{Properties of stacked images of the SXDS.  The table lists 
 pointing field names, final PSF sizes, total areas covered by the
 stacked images, and effective usable areas after excluding the 
masked regions in each pointing. Five images in different bands in each
 pointing have the same PSF size. The total and effective areas covered by the 5
 pointings taking account of overlapping areas between each pointing are
 4404.4 arcmin${}^2$ and 4057.0 arcmin${}^2$, respectively.}
\end{deluxetable}

\clearpage
\begin{deluxetable}{lcl}
\tablecaption{LIST OF PARAMETERS IN THE SXDS OPTICAL IMAGING OBJECT CATALOGS\label{tab:catalog}}
\tablehead{
\colhead{Parameter Name} & {Unit} & {Description}
}
\startdata
{ID}& {} & {sequential unique ID of the object in the catalog}\\
{X} & {pixels} & {X coordinate of detection}\\
{Y} & {pixels} & {Y coordinate of detection}\\
{RA} & {h:m:s} & {right ascension of detection}\\
{Dec} & {d:m:s} & {declination of detection}\\
{KRON\_RADIUS} & {pixels} & {Kron radius $r_K$ defined in Bertin \& Arnouts 1996}\\
{A\_IMAGE} & {pixels} & {2nd order moment along the major axis}\\
{B\_IMAGE} & {pixels} & {2nd order moment along the minor axis}\\
{ELLIPTICITY} & {} & {isophotal weighted ellipticity of the object (=A\_IMAGE/B\_IMAGE)}\\
{THETA\_IMAGE} & {degree} & {position angle of the major axis, counter-clockwised, 0.0=X axis}\\
{ISOAREA\_IMAGE} & {pixels} & {area of the lowest isophote}\\
{IsophotalMag(B)} & {mag} & {isophotal magnitude in B}\\
{IsophotalMag(V)} & {mag} & {isophotal magnitude in V}\\
{IsophotalMag(R)} & {mag} & {isophotal magnitude in Rc}\\
{IsophotalMag(i)} & {mag} & {isophotal magnitude in i'}\\
{IsophotalMag(z)} & {mag} & {isophotal magnitude in z'}\\
{IsophotCorMag(B)} & {mag} & {corrected isophotal magnitude in B defined by Bertin \& Arnouts 1996}\\
{IsophotCorMag(V)} & {mag} & {corrected isophotal magnitude in V defined by Bertin \& Arnouts 1996}\\
{IsophotCorMag(R)} & {mag} & {corrected isophotal magnitude in Rc defined by Bertin \& Arnouts 1996}\\
{IsophotCorMag(i)} & {mag} & {corrected isophotal magnitude in i' defined by Bertin \& Arnouts 1996}\\
{IsophotCorMag(z)} & {mag} & {corrected isophotal magnitude in z' defined by Bertin \& Arnouts 1996}\\
{2.0ApertureMag(B)} & {mag} & {$\phi\, 2''$ fixed-aperture magnitude in B}\\
{2.0ApertureMag(V)} & {mag} & {$\phi\, 2''$ fixed-aperture magnitude in V}\\
{2.0ApertureMag(R)} & {mag} & {$\phi\, 2''$ fixed-aperture magnitude in Rc}\\
{2.0ApertureMag(i)} & {mag} & {$\phi\, 2''$ fixed-aperture magnitude in i'}\\
{2.0ApertureMag(z)} & {mag} & {$\phi\, 2''$ fixed-aperture magnitude in z'}\\
{3.0ApertureMag(B)} & {mag} & {$\phi\, 3''$ fixed-aperture magnitude in B}\\
{3.0ApertureMag(V)} & {mag} & {$\phi\, 3''$ fixed-aperture magnitude in V}\\
{3.0ApertureMag(R)} & {mag} & {$\phi\, 3''$ fixed-aperture magnitude in Rc}\\
{3.0ApertureMag(i)} & {mag} & {$\phi\, 3''$ fixed-aperture magnitude in i'}\\
{3.0ApertureMag(z)} & {mag} & {$\phi\, 3''$ fixed-aperture magnitude in z'}\\
{MAG\_AUTO(B)} & {mag} & {automatic-aperture magnitude in B}\\
{MAG\_AUTO(V)} & {mag} & {automatic-aperture magnitude in V}\\
{MAG\_AUTO(R)} & {mag} & {automatic-aperture magnitude in Rc}\\
{MAG\_AUTO(i)} & {mag} & {automatic-aperture magnitude in i'}\\
{MAG\_AUTO(z)} & {mag} & {automatic-aperture magnitude in z'}\\
{MAG\_BEST(B)} & {mag} & {MAG\_AUTO(B) if there are no neighbors, or IsophotCorMag(B) otherwise}\\
{MAG\_BEST(V)} & {mag} & {MAG\_AUTO(V) if there are no neighbors, or IsophotCorMag(V) otherwise}\\
{MAG\_BEST(R)} & {mag} & {MAG\_AUTO(Rc) if there are no neighbors, or IsophotCorMag(R) otherwise}\\
{MAG\_BEST(i)} & {mag} & {MAG\_AUTO(i') if there are no neighbors, or IsophotCorMag(i) otherwise}\\
{MAG\_BEST(z)} & {mag} & {MAG\_AUTO(z') if there are no neighbors, or IsophotCorMag(z) otherwise}\\
{Err(IsophotalMag)(B)} & {mag} & {isophotal magnitude rms error in B}\\
{Err(IsophotalMag)(V)} & {mag} & {isophotal magnitude rms error in V}\\
{Err(IsophotalMag)(R)} & {mag} & {isophotal magnitude rms error in Rc}\\
{Err(IsophotalMag)(i)} & {mag} & {isophotal magnitude rms error in i'}\\
{Err(IsophotalMag)(z)} & {mag} & {isophotal magnitude rms error in z'}\\
{Err(IsophotCorMag)(B)} & {mag} & {corrected isophotal magnitude rms error in B}\\
{Err(IsophotCorMag)(V)} & {mag} & {corrected isophotal magnitude rms error in V}\\
{Err(IsophotCorMag)(R)} & {mag} & {corrected isophotal magnitude rms error in Rc}\\
{Err(IsophotCorMag)(i)} & {mag} & {corrected isophotal magnitude rms error in i'}\\
{Err(IsophotCorMag)(z)} & {mag} & {corrected isophotal magnitude rms error in z'}\\
{Err(2.0ApertureMag)(B)} & {mag} & {$\phi 2''$ fixed-aperture magnitude rms error in B}\\
{Err(2.0ApertureMag)(V)}  & {mag} & {$\phi 2''$ fixed-aperture magnitude rms error in V}\\
{Err(2.0ApertureMag)(R)}  & {mag} & {$\phi 2''$ fixed-aperture magnitude rms error in Rc}\\
{Err(2.0ApertureMag)(i)}  & {mag} & {$\phi 2''$ fixed-aperture magnitude rms error in i'}\\
{Err(2.0ApertureMag)(z)}  & {mag} & {$\phi 2''$ fixed-aperture magnitude rms error in z'}\\
{Err(3.0ApertureMag)(B)}  & {mag} & {$\phi 3''$ fixed-aperture magnitude rms error in B}\\
{Err(3.0ApertureMag)(V)}  & {mag} & {$\phi 3''$ fixed-aperture magnitude rms error in V}\\
{Err(3.0ApertureMag)(R)}  & {mag} & {$\phi 3''$ fixed-aperture magnitude rms error in Rc}\\
{Err(3.0ApertureMag)(i)}  & {mag} & {$\phi 3''$ fixed-aperture magnitude rms error in i'}\\
{Err(3.0ApertureMag)(z)}  & {mag} & {$\phi 3''$ fixed-aperture magnitude rms error in z'}\\
{Err(MAG\_AUTO)(B)}  & {mag} & {automatic-aperture magnitude rms error in B}\\
{Err(MAG\_AUTO)(V)}  & {mag} & {automatic-aperture magnitude rms error in V}\\
{Err(MAG\_AUTO)(R)}  & {mag} & {automatic-aperture magnitude rms error in Rc}\\
{Err(MAG\_AUTO)(i)}  & {mag} & {automatic-aperture magnitude rms error in i'}\\
{Err(MAG\_AUTO)(z)}  & {mag} & {automatic-aperture magnitude rms error in z'}\\
{Err(MAG\_BEST)(B)}  & {mag} & {MAG\_BEST magnitude rms error in B}\\
{Err(MAG\_BEST)(V)}  & {mag} & {MAG\_BEST magnitude rms error in V}\\
{Err(MAG\_BEST)(R)}  & {mag} & {MAG\_BEST magnitude rms error in Rc}\\
{Err(MAG\_BEST)(i)}  & {mag} & {MAG\_BEST magnitude rms error in i'}\\
{Err(MAG\_BEST)(z)}  & {mag} & {MAG\_BEST magnitude rms error in z'}\\
{FLUX\_MAX(B)} & {ADU} & {peak surface brightness in B}\\
{FLUX\_MAX(V)} & {ADU} & {peak surface brightness in V}\\
{FLUX\_MAX(R)} & {ADU} & {peak surface brightness in Rc}\\
{FLUX\_MAX(i)} & {ADU} & {peak surface brightness in i'}\\
{FLUX\_MAX(z)} & {ADU} & {peak surface brightness in z'}\\
{FWHM\_IMAGE(B)} & {pixels} & {FWHM of profile from a gaussian fit in B}\\
{FWHM\_IMAGE(V)} & {pixels} & {FWHM of profile from a gaussian fit in V}\\
{FWHM\_IMAGE(R)} & {pixels} & {FWHM of profile from a gaussian fit in Rc}\\
{FWHM\_IMAGE(i)} & {pixels} & {FWHM of profile from a gaussian fit in i'}\\
{FWHM\_IMAGE(z)} & {pixels} & {FWHM of profile from a gaussian fit in z'}\\
{CLASS\_STAR(B)} & {} & {stellarity index in B: 0.0=Galaxy and 1.0=Star}\\
{CLASS\_STAR(V)} & {} & {stellarity index in V: 0.0=Galaxy and 1.0=Star}\\
{CLASS\_STAR(R)} & {} & {stellarity index in R: 0.0=Galaxy and 1.0=Star}\\
{CLASS\_STAR(i)} & {} & {stellarity index in i': 0.0=Galaxy and 1.0=Star}\\
{CLASS\_STAR(z)} & {} & {stellarity index in z': 0.0=Galaxy and 1.0=Star}\\
{FLAGS(B)} & {} & {extraction flags in B defined by SExtractor}\\
{FLAGS(V)} & {} & {extraction flags in V defined by SExtractor}\\
{FLAGS(R)} & {} & {extraction flags in Rc defined by SExtractor}\\
{FLAGS(i)} & {} & {extraction flags in i' defined by SExtractor}\\
{FLAGS(z)} & {} & {extraction flags in z' defined by SExtractor}\\
{FLAG\_AREA}  & {} & {index for affected area: 0=Clean, 1=STRONG\_HALO, 2=WEAK\_HALO}\\
\enddata
\tablecomments{The list of fields listed in the optical
 multi-waveband catalogs. All the fields are commonly included 
in all the catalogs created for each detection and each pointing.}
\end{deluxetable}

\clearpage
\begin{deluxetable}{lc}
\tablecaption{THE SEXTRACTOR PARAMETERS USED TO CREATE THE SXDS CATALOGS\label{tab:sexparam}}
\tablehead{
 \colhead{Parameter} & 
 \colhead{Value} 
}
\startdata
DETECT\_MINAREA & 5 \\
DETECT\_THRESH & 2.0 \\
ANALYSIS\_THRESH & 2.0 \\
FILTER  & N \\
DEBLEND\_NTHRESH & 32 \\
DEBLEND\_MINCONT  & 0.005 \\
BACK\_SIZE  & 32 \\
BACK\_FILTERSIZE & 3 \\
BACKPHOTO\_TYPE & GLOBAL \\
BACKPHOTO\_THICK & 24 \\
SATUR\_LEVEL & 50000.0 \\
GAIN & 2.6 \\
\enddata
\tablecomments{Parameters for SExtractor used to create catalogs. The 
 above values are common to all the SExtractor runs regardless of
 detection bandpasses.}
\end{deluxetable}

\clearpage
\begin{deluxetable}{cccc}
\tablecaption{RESULT OF ASTROMETRIC CALIBRATION\label{tab:astmt}}
\tablehead{
 \colhead{} & \colhead{rms (RA)} &  \colhead{rms (Dec)} & \colhead{} \\
 \colhead{Pointing} & \colhead{(arcsec)} &  \colhead{(arcsec)} & \colhead{Number of Stars} 
}
\startdata
Center & 0.160 & 0.151 & 233 \\
North & 0.220 & 0.218  & 208 \\
South & 0.251 & 0.207 & 194 \\
East  & 0.261 & 0.225 & 199 \\
West  & 0.217 & 0.201 & 218 \\
\enddata
\tablecomments{The resultant r.m.s. of astrometric calibration. The
 number of stars used for calculation of the world coordinates is shown 
 in the far right column.}
\end{deluxetable}

\clearpage
\begin{deluxetable}{ccccccc}
\tablecaption{THE GALACTIC EXTINCTION IN THE SXDF\label{tab:gal_ext}}
\tablehead{
 \colhead{Pointing} & 
 \colhead{$E(B-V)$} & 
 \colhead{$A_B$} & 
 \colhead{$A_V$} & 
 \colhead{$A_{Rc}$} & 
 \colhead{$A_{i'}$} & 
 \colhead{$A_{z'}$} 
}
\startdata
{\it SXDS-C} & 0.021 & 0.091 & 0.070 & 0.056 & 0.044 & 0.031 \\
{\it SXDS-N} & 0.019 & 0.083 & 0.064 & 0.052 & 0.040 & 0.029 \\
{\it SXDS-S} & 0.023 & 0.100 & 0.077 & 0.062 & 0.048 & 0.034 \\
 {\it SXDS-E} & 0.021 & 0.091 & 0.070 & 0.056 & 0.044 & 0.031 \\
{\it SXDS-W} & 0.021 & 0.091 & 0.070 & 0.057 & 0.044 & 0.031 \\
\enddata
\tablecomments{The magnitude attenuation due to the Galactic extinction in
 each band calculated for the central coordinates of each pointing based 
 on Schlegel et al. (1998). Since magnitudes in the released catalogs are not corrected for the Galactic extinction, the above values must be taken into account when studying extragalactic objects. 
}
\end{deluxetable}
\clearpage
\begin{deluxetable}{ccccccccccc}
\tablecaption{THE SYSTEMATIC DIFFERENCES IN APERTURE MAGNITUDES BETWEEN
 THE DIFFERENT CATALOGS OF THE SXDS\label{tab:diff_magap}}
\tablehead{
 \colhead{Pointing} & 
 \colhead{$B(2'')$} & 
 \colhead{$B(3'')$} & 
 \colhead{$V(2'')$} & 
 \colhead{$V(3'')$} & 
 \colhead{$R_c(2'')$} &
 \colhead{$R_c(3'')$} &
 \colhead{$i'(2'')$} & 
 \colhead{$i'(3'')$} &
 \colhead{$z'(2'')$} &
 \colhead{$z'(3'')$}
}
\startdata
{\it C-N} & -0.04 & -0.02 & -0.04 & -0.01 & -0.03 & -0.01 & +0.03 & +0.02 & -0.02 & -0.01 \\
{\it C-S} & -0.05 & -0.02 & -0.02 & -0.00 & -0.03 &  0.00 & +0.03 & +0.00 & +0.00 & +0.01 \\
{\it C-E} & -0.02 & -0.01 & -0.02 & -0.01 & -0.01 & -0.00 & +0.05 & +0.01 & -0.01 & -0.00 \\
{\it C-W} & +0.01 & +0.01 & -0.03 & -0.01 & +0.01 & +0.00 & -0.01 & -0.00 & -0.01 & -0.01 \\
\enddata
\tablecomments{The systematic differences in aperture magnitudes of
 objects between the catalog of {\it SXDS-C} and that of another surrounding pointing. For
 each band, the magnitude differences in $2''$- and $3''$-diameter aperture magnitude are
 listed. The column of pointing denotes the pair of pointings
 compared. For instance, the line for {\it C-N} indicates magnitude
 differences calculated by magnitudes in {\it SXDS-C} minus those in {\it SXDS-N}. }
\end{deluxetable}

\clearpage
\begin{deluxetable}{cccc}
\tablecaption{CRITERIA USED FOR STAR/GALAXY SEPARATION\label{tab:sgsep}}
\tablehead{
 \colhead{Band} & 
 \colhead{Criteria}
}
\startdata
$B$ & $(m=19.5-23.0, \sigma>4.5, \sigma>-2.5\,(m-20.3)+4.5)$ or $(m>23.0)$ \\
$V$ & $(m=19.5-23.0, \sigma>4.5, \sigma>-2.5\,(m-20.0)+4.5)$ or $(m>23.0)$ \\
$R_c$ & $(m=19.5-22.5, \sigma>4.5, \sigma>-2.5\,(m-20.3)+4.5)$ or $(m>22.5)$ \\
$i'$ & $(m=19.5-22.5, \sigma>4.5, \sigma>-2.5\,(m-20.1)+4.5)$ or $(m>22.5)$ \\
$z'$ & $(m=18.5-22.5, \sigma>4.5, \sigma>-3.0\,(m-19.3)+4.5)$ or $(m>22.5)$ \\
\enddata
\tablecomments{The criteria used in star/galaxy separation procedure. 
Objects satisfying these criteria are recognized as galaxies. 
In the criteria list, $m$ denotes 2''-diameter fixed aperture
 magnitude and $\sigma$ is FWHM of objects in the corresponding band. }
\end{deluxetable}


\begin{thebibliography}{}
\bibitem[Arnouts et al. (1999)]{Arnouts99}
 Arnouts, S., D'Odorico, S., Cristiani, S., Zaggia, S., Fontana, A., \&
			       Giallongo, E. 1999, A\&A, {\bf 341}, 641

\bibitem[Arnouts et al. (2001)]{Arnouts01}
 Arnouts, S. et al. 2001, A\&A, {\bf 379}, 740

\bibitem[Bertin \& Arnouts (1996)]{Bertin96}
 Bertin, E. \& Arnouts, S. 1996, A\&AS, {\bf 117}, 393

\bibitem[Bolzonella et al. (2000)]{Bolzonella00}
 Bolzonella, M., Miralles, J.-M., Pell\'{o}, R. 2000, A\&A, {\bf 363}, 476

\bibitem[Capak et al. (2004)]{Capak04}
 Capak, P. et al. 2004, AJ, {\bf 127}, 180

\bibitem[Capak et al. (2007)]{Capak07}
 Capak, P. et al. 2007, \apjs, {\bf 172}, 99

\bibitem[Coppin et al. (2006)]{Copping06}
 Coppin, K. et al. 2006, MNRAS, {\bf 372}, 1621

\bibitem[Cutri et al. (2003)]{Cutri03)}
 Cutri, R. M. et al. 2003, VizieR On-line Data Catalog: II/246 (Originally published in University of Massachusetts and Infrared Processing and Analysis Center (IPAC/California Institute of Technology))

\bibitem[Dye et al. (2006)]{Dye06}
 Dye, S. et al. 2006, \mnras, {\bf 372}, 1227

\bibitem[Doi (2003)]{Doi03}
 Doi, M. et al. 2003, IAU Circ., {\bf 8119}, 1

\bibitem[Feldmann et al. (2006)]{Feldmann06}
 Feldmann, R. et al. 2006, \mnras, {\bf 372}, 565

\bibitem[Foucaud et al. (2007)]{Foucaud07}
 Foucaud, S. et al. 2007, \mnras, {\bf 376}, L20

\bibitem[Furusawa et al. (2000)]{Furusawa00}
 Furusawa, H., Shimasaku, K., Doi, M., \& Okamura, S. 2000, \apj, {\bf 534}, 624

\bibitem[Giavalisco et al. (2004)]{Giavalisco04}
 Giavalisco, M. et al. 2004, \apj, {\bf 600}, L93

\bibitem[Gunn \& Stryker (1983)]{Gunn83}
 Gunn, J. E. \& Stryker, L. L. 1983, \apjs, {\bf 52}, 121

\bibitem[Hogg et al. (1997)]{Hogg97}
 Hogg, D. W., Pahre, M. A., McCarthy, J. K., Cohen, J. G., Blandford, R., Smail, I., Soifer, B. T. 1997, \mnras, {\bf 288}, 404

\bibitem[Huan et al. (2001)]{Huang01}
 Huang, J.-S. et al. 2001, A\&A, {\bf 368}, 787

\bibitem[Kashikawa et al. (2004)]{Kashikawa04}
 Kashikawa, N. et al. 2004, PASJ, {\bf 56}, 1011

\bibitem[Kodama et al. (2004)]{Kodama04}
 Kodama, T. et al. 2004, \mnras, {\bf 350}, 1005

\bibitem[Kron (1980)]{Kron80}
 Kron, R. G. 1980, \apjs, {\bf 43}, 305

\bibitem[Kummel \& Wagner (2001)]{Kummel01}
 K\"{u}mmel, M. W. \& Wagner, S. J. 2001, A\&A, {\bf 370}, 384

\bibitem[Lawrence et al. (2007)]{Lawrence07}
 Lawrence, A. et al. 2007, \mnras, {\bf 379}, 1599

\bibitem[Lidman et al. (2005)]{Lidman2005}
 Lidman, C. et al. 2005, A\&A, {\bf 430}, 843

\bibitem[Lilly et al. (1991)]{Lilly91}
 Lilly, S. J., Cowie, L. L., \& Gardner, J. P. 1991, \apj, {\bf 369}, 79
		      
\bibitem[Lonsdale et al. (2003)]{Lonsdale03}
 Lonsdale, C. J. et al. 2003, \pasp, {\bf 115}, 897

\bibitem[Metcalfe et al. (2001)]{Metcalfe01}
 Metcalfe, N., Shanks, T., Campos, A., McCracken, H. J., \& Fong, R. 2001, \mnras, {\bf 323}, 795

\bibitem[Miyazaki et al. (2002)]{Miyazaki04}
 Miyazaki, S. et al. 2002, PASJ, {\bf 54}, 833

\bibitem[Mortier et al. (2005)]{Mortier05}
 Mortier, A. M. J. et al. 2005, \mnras, {\bf 363}, 563

\bibitem[Ouchi et al. (2001)]{Ouchi01}
 Ouchi, M. et al. 2001, \apj, {\bf 558}, L83

\bibitem[Ouchi (2003)]{Ouchi03}
 Ouchi, M. 2003, PhD thesis, University of Tokyo

\bibitem[Ouchi et al. (2004)]{Ouchi04}
 Ouchi, M. et al. 2004, \apj, {\bf 611}, 660

\bibitem[Ouchi et al. (2005)]{Ouchi05}
 Ouchi, M. et al. 2005, \apj, {\bf 620}, L1

\bibitem[Ouchi et al.(2007)]{2007arXiv0707.3161O} 
 Ouchi, M. et al. 2007, ArXiv e-prints, 707, arXiv:0707.3161
%

\bibitem[Postman et al. (1998)]{Postman98}
 Postman, M., Lauer, T. R., Szapudi, I., \& Oegerle, W. 1998, \apj, {\bf 506}, 33 

\bibitem[Schlegel et al. (1998)]{Schlegel98}
 Schlegel, D. J., Finkbeiner, D. P., \& Davis, M. 1998, \apj, {\bf 500}, 525

\bibitem[Scoville et al. (2007)]{Scoville07}
 Scoville, N. et al. 2007, \apjs, {\bf 172}, 38

\bibitem[Simpson et al. (2006)]{Simpson06}
 Simpson, C. et al. 2006, \mnras, {\bf 372}, 741

\bibitem[Smail et al. (1995)]{Smail95}
 Smail, I, Hogg, D. W., Yan, L., \& Cohen, J. G. 1995, \apj, {\bf 449}, L105

\bibitem[Steidel \& Hamilton (1993)]{Steidel93}
 Steidel, C. C. \& Hamilton, D. 1993, AJ, {\bf 105}, 2017

\bibitem[Tyson (1988)]{Tyson88}
 Tyson, J. A. 1988, AJ, {\bf 96}, 1

\bibitem[Ueda et al. (2007)]{Ueda07}
 Ueda, Y. et al. 2007, \apjs, submitted

\bibitem[Yagi et al. (2002)]{Yagi02}
 Yagi, M., Kashikawa, N., Sekiguchi, M., Doi, M., Yasuda, N., Shimasaku, K., \& Okamura, S. 2002, \aj, {\bf 123}, 66

\bibitem[Yamada et al. (2005)]{key-10}
 Yamada, T. et al. 2005, \apj, {\bf 634}, 861

\bibitem[Yasuda et al. (2001)]{key-10}
 Yasuda, N. et al. 2001, AJ, {\bf 122}, 1104

\end{thebibliography}
\end{document}